\documentclass[%
 aip,
 amsmath,amssymb,
 reprint,%
]{revtex4-1}

\usepackage{graphicx}%
\usepackage{dcolumn}%
\usepackage{bm}%

\usepackage[utf8]{inputenc}
\usepackage[T1]{fontenc}
\usepackage{mathptmx}
\usepackage{etoolbox}
\usepackage{color}
\usepackage{SIunits}

\makeatletter
\def\@email#1#2{%
 \endgroup
 \patchcmd{\titleblock@produce}
  {\frontmatter@RRAPformat}
  {\frontmatter@RRAPformat{\produce@RRAP{*#1\href{mailto:#2}{#2}}}\frontmatter@RRAPformat}
  {}{}
}%
\makeatother
\begin{document}

\newcommand{\gb}{\textcolor{red} }
\newcommand{\tf}{\textcolor{blue} }
\preprint{AIP/123-QED}

\title{Refinement of molecular dynamics ensembles using experimental data and flexible forward models}
\author{Thorben Fr\"ohlking}%
\altaffiliation[Current address:]{ Universit\'e{} de Gen\`eve, Switzerland}%
\author{Mattia Bernetti}%
\altaffiliation[Current address:]{ Universit\`a{} di Bologna, Italy}%
\author{Giovanni Bussi}
\email{bussi@sissa.it}
\affiliation{ 
Scuola Internazionale Superiore di Studi Avanzati, via Bonomea 265, 34136 Trieste, Italy
}%

\date{\today}%

\begin{abstract}
A novel method combining maximum entropy principle, the Bayesian-inference of ensembles approach,
and the optimization of empirical forward models is presented.
Here we focus on the Karplus parameters for RNA systems, which relate the dihedral angles of $\gamma$, $\beta$, and the dihedrals in the sugar ring to the corresponding $^3J$-coupling signal between coupling protons.
Extensive molecular simulations are performed on a set of RNA tetramers and hexamers and combined with available nucleic-magnetic-resonance data.
Within the new framework, the sampled structural dynamics can be reweighted to match experimental data while the error arising from inaccuracies in the forward models can be corrected simultaneously and consequently does not leak into the reweighted ensemble.
Carefully crafted cross-validation procedure and regularization terms enable obtaining transferable Karplus parameters.
Our approach identifies the optimal regularization strength and new sets of Karplus parameters balancing good agreement between simulations and experiments with minimal changes to the original ensemble.
\end{abstract}

\maketitle

\begin{quotation}
Abstract.
\end{quotation}

Molecular dynamics (MD) simulations are a fundamental tool to model the conformational dynamics of molecular systems \cite{hollingsworth2018molecular}.
In order to improve their accuracy, it is more and more common to use integrative approaches where MD simulations are combined
with experimental data \cite{bonomi2017principles,bottaro2018biophysical,bernetti2023integrating}. This makes it possible to either improve the employed force fields \cite{norgaard2008experimental,li2011iterative,wang2012systematic,wang2014building,cesari2016combining,cesari2019fitting,frohlking2020toward,kofinger2021empirical,frohlking2022automatic} or to directly refine the generated ensembles.
In particular, ensembles can be constructed by enforcing agreement with the experiment on the fly
\cite{cavalli2013molecular,white2014efficient,hummer2015bayesian,bonomi2016metainference,cesari2016combining}
or by refining
the ensembles \emph{a posteriori} \cite{costa2022reweighting}, using either selection \cite{bernado2007structural,tria2015advanced} or reweighting
\cite{pitera2012use,hummer2015bayesian,brookes2016experimental,kofinger2019efficient,bottaro2020integrating,medeiros2021comparison}  approaches. Methods based on the idea of minimally perturbing
the initially generated ensemble
are particularly appealing as they maximally use the microscopic information generated by the MD simulation
and only modify it when deviation with respect to the experiment is observed \cite{pitera2012use,cesari2018using}.
All the mentioned methods are based on a fundamental step, namely the back-calculation of the experimental
data from a simulated ensemble. The formulas used for this aim are often referred to as \emph{forward models}.
Forward models based on pairwise distances or angles between atoms or bonds are for instance used to back-calculate nuclear-magnetic-resonance (NMR) data,
whereas more complex calculations involving sums over all the observed atoms and possible solvent effects are used to back-calculate small-angle X-ray
scattering \cite{svergun1995crysol,kofinger2013atomic,knight2015waxsis} or cryo-electron-microscopy data \cite{bonomi2019bayesian}.
Many  forward models are based on empirical relationships, as is the case for the so-called Karplus equations \cite{karplus1963vicinal}.
However, the parameters of these relationships are usually determined once and for all and are not optimized for the specific problem.

In this paper, we propose a procedure to systematically refine parametrized forward models by combining MD simulations performed
on multiple systems. The refinement is simultaneously done on the simulated ensembles and on the forward models so that
it is not necessary to have an accurate reference ensemble to start with. The procedure is here applied to the refinement of
Karplus equations used to back-calculate $^3J$ scalar couplings in RNA systems, by using extensive MD simulations of a number 
of RNA oligomers for which NMR experimental data are available. The formalism is here derived
starting from the Bayesian-inference of ensembles method \cite{hummer2015bayesian}
but can be also seen as an extension of the equivalent regularized maximum entropy  approach \cite{cesari2018using}.

\section{Methods}

\subsection{The Bayesian-inference of ensembles method}
We consider a molecular system whose conformation is denoted by the high-dimensional vector $\mathbf{x}$.
The canonical distribution function is
$\rho_0(\mathbf{x})\propto e^{-\frac{U(\mathbf{x})}{k_BT}}$,
where $U(\mathbf{x})$ is the energy calculated using a molecular force field,
$k_B$ is the Boltzmann constant, and $T$ the temperature.
We use as a starting point the Bayesian inference of ensemble (BioEn) method \cite{hummer2015bayesian},
which requires minimizing the following cost function
\begin{equation}
\label{eq:cost-bioen}
\mathcal{L}_{\textrm{BioEn}}[\rho] = \frac{\chi^2[\rho]}{2} - \tilde{\alpha} S[\rho|\rho_0]~,
\end{equation}
thus obtaining the refined distribution $\rho(\mathbf{x})$.

The first term in Eq.~\ref{eq:cost-bioen} contains $\chi^2$, which
represents the discrepancy
between simulation and experiment and is defined as
\begin{equation}
\label{eq:def-chi2}
\chi^2[\rho]=\sum_i^{N_{\textrm{exp}}}
\frac{
\left(
\bar{g}_i[\rho] - g_{i,\textrm{exp}}
\right)^2
}{
\sigma^2_i
}~.
\end{equation}
The variable $N_{\textrm{exp}}$ represents the total number of experimental data points that are available. These data points can come from different sources, including multiple measurements taken in one experiment (such as in small-angle X-ray scattering), multiple separate experiments, or a combination of both.
$g_{i,\textrm{exp}}$ is the value of the experimental data,
$\sigma_i$ its associated uncertainty, and
$\bar{g}_i$ the value back-calculated from the refined ensemble, which is computed as
\begin{equation}
\bar{g}_i[\rho] = \int d\mathbf{x} \rho(\mathbf{x}) g_i(\mathbf{x})~.
\end{equation}
The function used to back-calculate the experimental values ($g_i(\mathbf{x})$)
can have an arbitrary functional form.  For instance, for nuclear Overhauser effect (NOE) experiments,
it is the inverse of the sixth power of the distance between the involved atoms and,
for $^3J$ scalar couplings, it is an empirically parametrized Karplus equation.

The second term in Eq.~\ref{eq:cost-bioen}
contains a hyperparameter ($\tilde{\alpha}>0$, named $\theta$ in the original BioEn method) that controls the relative confidence that we have in the initial ensemble and in the experimental data, and multiplies the relative entropy
between the refined ensemble ($\rho$) and the original ensemble ($\rho_0$),
defined as 
\begin{equation}
S[\rho|\rho_0] = -\int d\mathbf{x} \rho(\mathbf{x}) \ln \frac{\rho(\mathbf{x})}{\rho_0(\mathbf{x})}~.
\end{equation}
In the limit of small (positive) $\tilde{\alpha}$, the minimization is dominated by $\frac{\chi^{2}}{2}$. If one assumes that one or more ensembles that agree with all experimental data points exist, the minimum $\chi^{2}$ will be exactly equal to zero. Among all the ensembles that agree with the experiment, the method will choose the one that has the maximum relative entropy. In this regime, this approach is thus exactly equivalent to the maximum entropy principle \cite{pitera2012use}.

The form of $\rho$ that maximizes the relative entropy at a fixed value of the back-calculated observables ($\bar{g}_i[\rho]$) can be shown to be 
\cite{pitera2012use,cesari2018using}:
\begin{equation}
\rho_{\lambda}(\mathbf{x})\propto\exp\left(-\sum_{i=1}^{N_{exp}}\lambda_{i}g_{i}(\mathbf{x})\right).
\label{eq:rho-of-lambda}
\end{equation}
This allows rephrasing the dependence of $\mathcal{L}$ on $\rho$ as a dependence on the free parameters $\lambda_i$, one for each
experimental data point. It is also common to introduce a function
\begin{multline}
\Gamma(\lambda)=\ln\int d\mathbf{x}\rho(\mathbf{x}) P_{0}(\mathbf{x})e^{-\sum_{i}\lambda_{i}\left(g_{i}(\mathbf{x})-g_{i,exp}\right)}= \\S[P_{\lambda}|P_{0}]-\sum_{i}\lambda_{i}\left(\bar{g}_{i}[P_{\lambda}]-g_{i,exp}\right)~.
\label{eq:gamma}
\end{multline}
The gradient of this function is
\begin{equation}
\frac{\partial \Gamma}{\partial\lambda_{i}}=\left(g_{i,exp}-\bar{g}_{i}[P_{\lambda}]\right)~.
\end{equation}
Thus, if a set of $\{\lambda\}$ such that back-calculated observables are identical to the experimental ones exists,
$\Gamma(\lambda)$ will be minimal. In addition, since $\Gamma$ is convex, this will be a global minimum \cite{cesari2018using}.

 Finite values of $\tilde{\alpha}$ can be used to better regularize the fitting procedure, and can be shown to be equivalent to directly modeling experimental errors in the maximum-entropy framework \cite{cesari2016combining}. We notice that Eq.~\ref{eq:rho-of-lambda} is also valid for a finite $\tilde{\alpha}$. By exploiting the relationship between $\Gamma$ and $S$ (Eq.~\ref{eq:gamma}), one can rewrite the cost function as
 \begin{equation}
 \mathcal{L}_{\textrm{BioEn}}(\lambda)=\frac{\chi^{2}[\rho_{\lambda}]}{2}-\tilde{\alpha}\left(\Gamma(\lambda)+\sum_{i}\lambda_{i}\left(\bar{g}_{i}[\rho_{\lambda}]-g_{i,exp}\right)\right)
 \end{equation}
The gradient of this function with respect to $\lambda$ is
 \begin{equation}
\frac{\partial\mathcal{L}_{\textrm{BioEn}}(\lambda)}{\partial\lambda_{i}}=\sum_{j}\frac{\partial\bar{g}_{j}[\rho_{\lambda}]}{\partial\lambda_{i}}\left(\frac{\bar{g}_{j}[\rho_{\lambda}]-g_{j,exp}}{\sigma_{j}^{2}}-\tilde{\alpha}\lambda_{j}\right)
 \end{equation}
This gradient can be set to zero by setting $\bar{g}_{i}[\rho_{\lambda}]=g_{i,exp}+\tilde{\alpha}\sigma_{i}^{2}\lambda_{i}$. The interpretation is the following: for a finite $\tilde{\alpha}$, the back-calculated observables ($\bar{g}_{i}[\rho_{\lambda}]$) will not match exactly the experiment ($g_{i,exp}$). The larger $\tilde{\alpha}$ is, the larger a mismatch will be accepted.

Remarkably, the same condition can be obtained by minimizing a regularized version of the $\Gamma$ function above, defined as
\begin{equation}
\tilde{\Gamma}(\lambda)=\Gamma(\lambda)+\frac{\tilde{\alpha}\sum_{i}\sigma_{i}^{2}\lambda_{i}^{2}}{2}
\label{eq:gamma-tilde}
\end{equation}
In other words, a regularization proportional to the relative entropy in Eq.~\ref{eq:cost-bioen} is completely equivalent to a L2 regularization on $\Gamma$. By defining the optimal value of $\lambda$ that minimizes $\tilde{\Gamma}$ as:
\begin{equation}
\lambda^{*}=\arg\min_{\lambda}\left(\tilde{\Gamma}(\lambda)\right)
\end{equation}
and by exploiting the fact that
$\bar{g}_{i}[\rho_{\lambda^*}]=g_{i,exp}+\tilde{\alpha}\sigma_{i}^{2}\lambda_i^*$ it is possible to show that:
\begin{equation}
\mathcal{L}_{\textrm{}}(\lambda^{*}) = -\tilde{\alpha}\tilde{\Gamma}(\lambda^{*})
\label{eq:l-of-gamma}
\end{equation}
Notice that this identify is only valid for $\lambda = \lambda^*$, that is, after the function $\tilde{\Gamma}$ has been minimized.
Also notice that, for any finite choice of $\tilde{\alpha}$,  $\tilde{\Gamma}$ is strictly convex and has a single minimum.
The advantage of minimizing $\tilde{\Gamma}(\lambda)$, which only depends on $N_{exp}$ degrees of freedom, rather than minimizing
$\mathcal{L}[\rho]$, which has a functional dependence on $\rho$, has been already recognized in other works \cite{cesari2018using,bottaro2020integrating}.
To clarify, if the variable $\rho$ is defined as a collection of weights assigned to various snapshots in an analyzed trajectory, then the function $\mathcal{L}[\rho]$ will take an input argument whose dimensionality is equal to the number of snapshots. This number of snapshots is typically much greater than the number of experimental data points denoted by $N_{exp}$.

\subsection{Flexible forward models}

The functions $g_i(\mathbf{x})$ are usually given a priori. In this work, we consider the possibility to fine-tune these functions so as to maximize the agreement between simulation and experiment. To this aim, we define a new cost function as
\begin{equation}
\mathcal{L}[\rho,\theta]=\frac{\chi^{2}[P,\theta]}{2}-\tilde{\alpha} S[\rho|\rho_{0}]+\tilde{\beta} R(\theta|\theta_{0})~.
\label{eq:cost}
\end{equation}
Here $\theta$ are the parameters controlling the forward models, $R$ is a regularization term that penalizes too large deviations from the original forward-model parameters, and $\tilde{\beta}$ the associated hyperparameter.

To minimize this function, we first rewrite it as a function of the free parameter $\lambda_i$ and of the forward-model parameters $\theta$:
\begin{equation}
\mathcal{L}(\lambda,\theta)=\frac{\chi^{2}[\rho_{\lambda},\theta]}{2}-\tilde{\alpha} S[\rho_{\lambda}|\rho_{0}]+\tilde{\beta} R(\theta|\theta_{0})~.
\label{eq:cost2}
\end{equation}

For practical purposes, it is convenient to rephrase the minimization of $\mathcal{L}$ as a nested minimization over $\theta$ and $\lambda$. The minimization over $\lambda$ is done in an inner loop and can be performed as discussed in the previous section, namely minimizing the convex regularized function $\tilde{\Gamma}(\lambda)$ for a fixed choice of the forward-model parameters $\theta$. Since this function depends on $\theta$, we reference to it as $\tilde{\Gamma}_{\theta}(\lambda)$, which emphasizes that the maximum entropy method should be applied with a fixed set of forward-model parameters $\theta$. We define the solution of the inner minimization as:
\begin{equation}
\lambda_{\theta}^{*}=\arg\min_{\lambda}\left(\tilde{\Gamma}_{\theta}(\lambda)\right)~.
\end{equation}In the outer loop, one should miminize the function $\mathcal{L}(\theta)$, which, using Eq.~\ref{eq:l-of-gamma}, can be written as:
\begin{equation}
\mathcal{L}(\theta)=-\tilde{\alpha}\tilde{\Gamma}_{\theta}(\lambda_{\theta}^{*})+\tilde{\beta} R(\theta|\theta_{0})~.
\end{equation}
In order to compute the gradient of $\mathcal{L}(\theta)$ one can consider that $\tilde{\Gamma}$ is in a stationary point as a function of $\lambda$, so that it is not necessary to consider the derivative of $\lambda_{\theta}^{*}$ with respect to $\theta$. One thus obtains:
\begin{multline}
\label{eq:l-gradient-theta}
\frac{\partial\mathcal{L}(\theta)}{\partial\theta}=-\tilde{\alpha}\frac{\partial\tilde{\Gamma}_{\theta}(\lambda_{\theta}^{*})}{\partial\theta}+\tilde{\beta}\frac{\partial R(\theta|\theta_{0})}{\partial\theta}=\\ \tilde{\alpha}\sum_{i}\lambda_{i}\int d\mathbf{\mathbf{x}}P(\mathbf{x})\frac{\partial g_{\theta,i}(\mathbf{x})}{\partial\theta}+\tilde{\beta}\frac{\partial R(\theta|\theta_{0})}{\partial\theta}
\end{multline}

In short, we iteratively minimize $\mathcal{L}$ as a function of the forward-model parameters $\theta$ (outer minimization). For each iteration, we:
\begin{itemize}
  \item Minimize the function $\tilde{\Gamma}(\lambda)$ at fixed forward-model parameters $\theta$ (inner minimization).
  \item Compute the gradients of the cost function $\mathcal{L}$ with respect to $\theta$ using Eq.~\ref{eq:l-gradient-theta}.
\end{itemize}

Notice that the inner minimization is done for a function that is proportional to the negative of the function used in the outer minimization.
This means that the nested minimization exactly correspond to the following minimax problem:
\begin{equation}
\min_{\theta} \max_{\lambda} \left( - \tilde{\alpha} \tilde{\Gamma}_{\theta}(\lambda) \right)~.
\label{eq:minimax}
\end{equation}
The inner minimization is guaranteed to be convex. The outer minimization instead might have multiple solutions.
For simplicity, we used the  L-BFGS-B algorithm \cite{morales2011remark} for both minimizations
as implemented in the SciPy library \cite{virtanen2020scipy}.

\subsection{Calculations using reweighting}

In the above derivations we assumed to be able to compute averages in the form:
\begin{equation}
\int d\mathbf{x} \rho_{\theta,\lambda}(\mathbf{x}) f(\mathbf{x})
\end{equation}
with $f(\mathbf{x})=g_{\theta}(\mathbf{x})$, to compute the back-calculated observables,
or with $f(\mathbf{x})=\frac{\partial g_{\theta}(\mathbf{x})}{\partial \theta}$, to compute the gradients of the back-calculated observables
with respect to the forward-model parameters. In both cases, the integrals are here replaced with weighted averages performed over
a set of snapshots sampled from an initial simulated trajectory:
\begin{equation}
\int d\mathbf{x} \rho_{\theta,\lambda}(\mathbf{x}) f(\mathbf{x})
\approx
\frac{
\sum_t w_t f(\mathbf{x}_t)
}{
\sum_t w_t
}
\end{equation}
where the non-normalized weights $w$ are obtained as
\begin{equation}
w_t = e^{-\frac{\sum_i \lambda_i g_{i,\theta}(\mathbf{x}_i)}{k_BT}}~.
\end{equation}
This relationship can be straightforwardly adjusted to take into account initial weights if the trajectories have been generated
using enhanced sampling methods based on a biasing potential. This is not necessary for the application presented here.

\subsection{Generalization to multiple systems}

The formalism above can be generalized to simultaneously fit over multiple systems. This is an advantage when the same forward models can be expected to be valid for multiple data points across different systems.
A fit including multiple systems would thus result in more transferable forward models.
To do so, one can just average the contribution of $N_{sys}$ systems:
\begin{equation}
\mathcal{L}[\{\lambda\},\theta]=-\frac{1}{N_{sys}}\sum_{k=1}^{N_{sys}}\tilde{\alpha}\tilde{\Gamma}_{\theta,k}(\lambda_{\theta,k}^{*})+\tilde{\beta} R(\theta|\theta_{0})~.
\end{equation}
Here, each system has a separate set of $\lambda$ parameters, corresponding to each of the available experimental data points.
The forward-model parameters $\theta$ instead are shared across all the analyzed systems.

Notice that, in line of principle, each system could be regularized with a separate hyper-parameter $\tilde{\alpha}$. The hyper-parameter $\tilde{\alpha}$, indeed, reports on how accurate we expect the prior distribution to be. In addition, a separate weight might be used for the different systems, so as to encode how much we would like each system to be relevant in the fit. In this work, we consider a homogeneous set of systems so that we don't exploit this possibility, which however might be useful if one wants to mix more heterogeneous sets of experimental data.

We also notice that, especially when combining multiple systems, it is convenient to report the $\chi^2$ discrepancy as an average, rather than a sum, over
all datapoints and all systems. For clarity, we explicitly indicate this average as a reduced $\chi^2$ ($\chi^2_{\textrm{red}}$).

\subsection{Interpretation of the hyperparameters $\tilde{\alpha}$ and $\tilde{\beta}$}

The hyperparameter $\tilde{\alpha}$ tunes how much we consider the prior ensemble trustable with respect to the experimental data points. Similarly, the hyperparameter $\tilde{\beta}$ tunes how much the original forward models are considered reliable. One can consider the following limiting cases:

\begin{itemize}
\item
$\tilde{\alpha}\rightarrow\infty, \tilde{\beta}\rightarrow\infty$: in this case, both the ensemble and the forward models are not modified ($\rho=\rho_{0}$ and $\theta=\theta_{0}$). The resulting $\chi^{2}$ reports the mismatch between theory and experiment as it is before ensemble refinement.
\item
$\tilde{\alpha}\rightarrow\infty, \tilde{\beta}\rightarrow0$: in this case, the ensemble is considered as perfectly trustable and thus not modified ($\rho=\rho_{0}$), but the forward models are adjusted so as to minimize the mismatch between theory and experiment. We notice that typically the same forward model is used for multiple data points and not flexible enough to simultaneously fit all of them. Thus, this fitting typically results in $\chi^{2}>0$.
\item
$\tilde{\alpha}\rightarrow0,\tilde{\beta}\rightarrow\infty$: in this case the ensemble is maximally modified to match experimental data, without modifying the forward models. Unless one is in a situation where no ensemble can be constructed compatible with the experimental data, this would result in $\chi^{2}\rightarrow0$. This is the most standard application of the maximum entropy method, without accounting for experimental errors.
\item
$\tilde{\alpha}\rightarrow0, \tilde{\beta}\rightarrow0$: in this case, both the ensemble and the forward models are modified. Multiple solutions might be possible, where either the ensemble or the forward models are given a different relevance, depending on the order in which the two limits are taken.
\end{itemize}

Limiting cases where the hyperparameters tend to zero can lead to overfitting.
By choosing finite values for $\tilde{\alpha}$ and $\tilde{\beta}$, we can optimize the performance of the procedure.
In particular, the original ensemble can be modified to agree with the experimental data while accounting for errors both in the experiments and in the forward models used.

\subsection{Using cross-validation to fine-tune hyperparameters $\tilde{\alpha}$ and $\tilde{\beta}$}

\begin{figure*}
\includegraphics[width=0.5\textwidth]{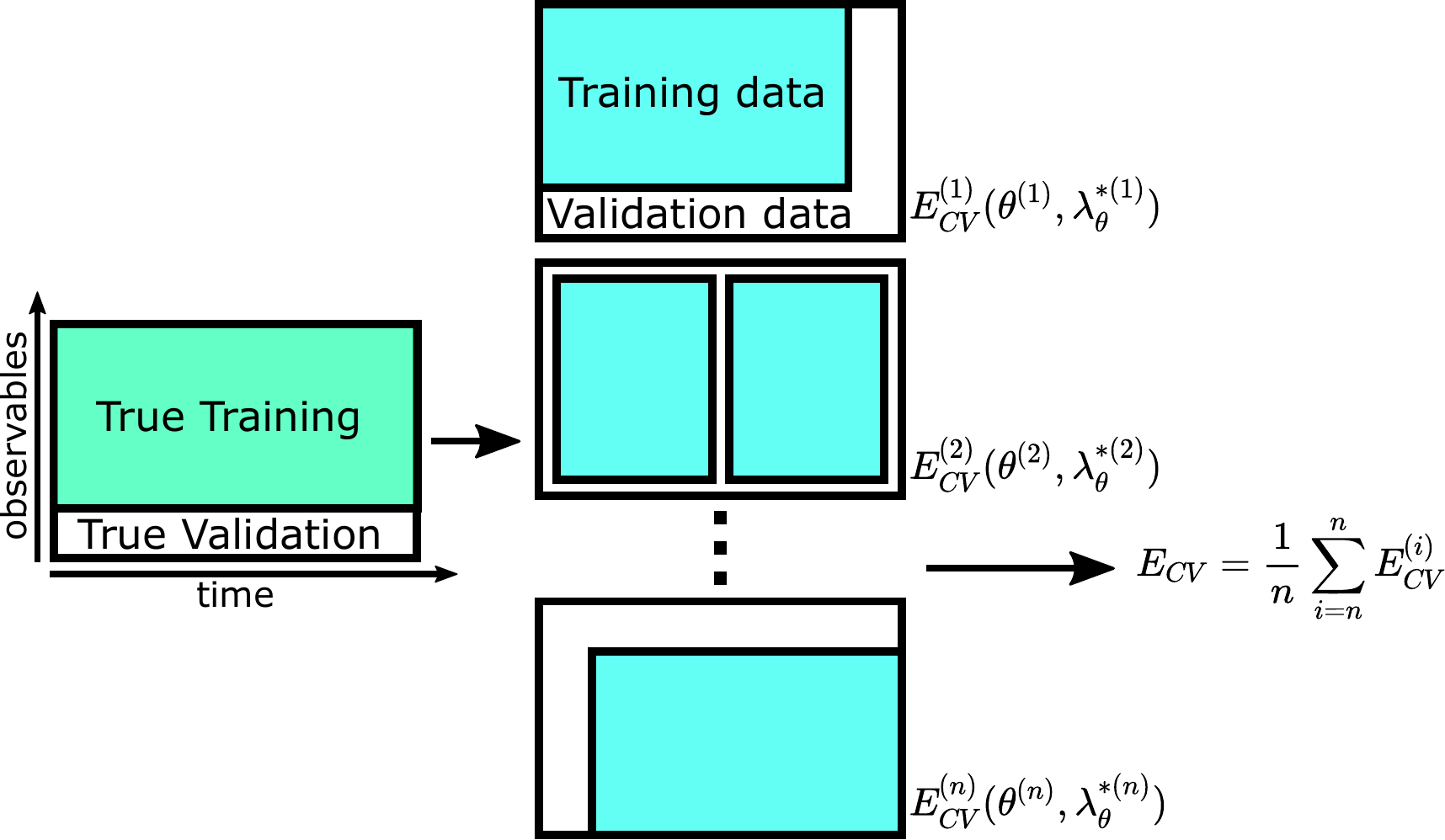}
\caption{\label{fig:SchematicCV}
Schematic depiction of the cross-validation procedure used in this study to decrease overfitting and allow more generalizable parameters. Datasets are shown as 2D objects, which represent the time in one dimension and space of available observables in the other. In order to keep data for validation after the fitting procedure including cross-validation and optimization on the training dataset at the optimal regularization strength has been performed, 20$\%$ of data were excluded from the fitting, here called 'True Validation'. Next the 'True Training' dataset is split $n$-times ($n=5$ in this study) randomly into blocks for training containing 70$\%$ of all the data and blocks for validation containing the remaining 30$\%$.
To minimize correlation between blocks, the trajectories are first split into 10 blocks along the dimension corresponding to time and subsequently we select a number of experimental data which result in a training or validation with 70$\%$ or 30$\%$ respectively.
The cross-validation error $E_{CV}^{(n)}$ is obtained by training the parameters $\theta^{(n)}$ and $\lambda^{*(n)}$ on the training data and evaluating their performance on the left-out validation block.
This is repeated once for each block, yielding multiple sets of trained parameters $(\theta^{(1)},\lambda^{*(1)}),\dots,(\theta^{(n)},\lambda^{*(n)})$ and the average over the $n$ results is the cross-validation error $E_{CV}$.}
\end{figure*}

Hyperparameters should be chosen based on the reliability
of (a) experimental data points, (b) reference forward models, and (b) original ensembles. As described in more detail in Ref.~\onlinecite{froehlking2020toward}, this
can be assessed by performing a cross-validation test where some of
the experiments are left out in training and tested in validation.
In $n$-fold cross validation, the dataset is randomly split into $n$ blocks of equal size.
One block at a time is left out of the training set and only used to compute the cross-validation error.
The parameters are trained on the remaining $n-1$ blocks.
This is repeated once for each block, yielding multiple sets of trained parameters $\lambda$ and $\theta$. The error obtained with these parameters in reproducing the left-out block is then computed, and the average over the $n$ results is the cross-validation error. Since $\tilde{\beta}$ mostly affects the fitting of the forward-model parameters $\theta$, it is crucial that in the cross-validation procedure the training and left-out
sets of experiment share the same forward models. In this way, it
is possible to see if the forward models trained on one set of experiments
are usable on a separate set of experiments. In addition, since the
evaluation of $\chi^{2}$ is affected by statistical errors and might
thus differ when using different time windows of the same simulation,
cross-validation tests should be designed to also validate parameters
trained on one simulation window against a different validation window \cite{frohlking2022automatic}.
In Fig.~\ref{fig:SchematicCV} we show how errors were computed and the data subdivided.
In practice, performing a cross validation using a cost function in the form of Eq.~\ref{eq:cost} has an important practical limitation: modifications
of the original forward-model parameters in regions that were not explored in the analyzed simulations will not enter in any way in the cross-validation
cost function. As discussed in the Results section, by simply choosing the value of $\tilde{\beta}$ that minimizes the cost function calculated on the cross-validation set
results in a not-sufficiently regularized fit, that will alter the forward models too much. For this reason, we used the cross-validation cost-function to
pick the optimal $\tilde{\alpha}$ only. The optimal $\tilde{\beta}$ was instead chosen to also ensure that the optimized forward models are not excessively
modified with respect to the original ones.

\subsection{Simulation details}

\begin{figure*}
\includegraphics[width=0.5\textwidth]{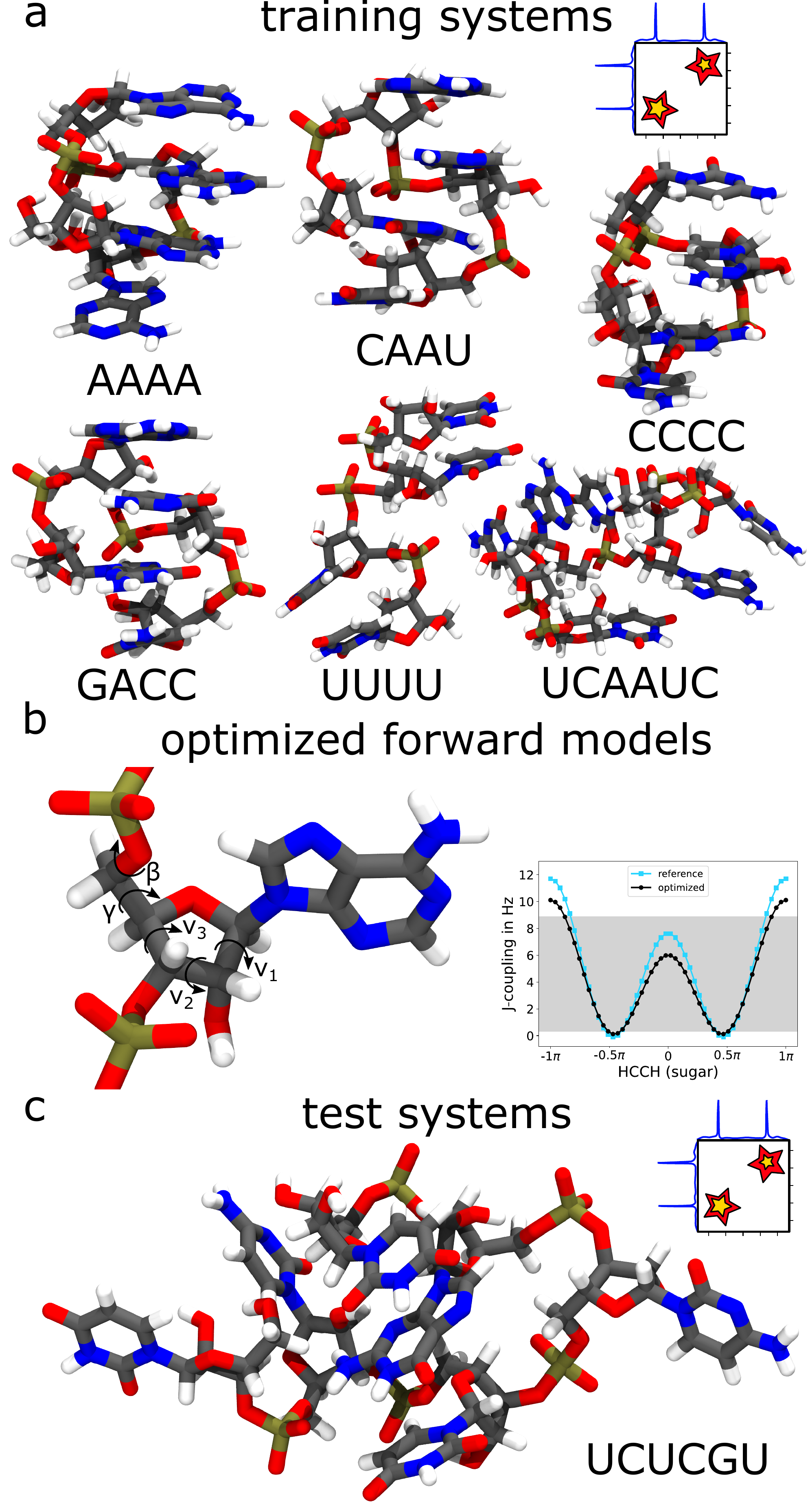}
\caption{\label{fig:SchematicKarplus}
(a) Systems included in the training set. The conformational ensembles of these systems are refined simultaneously with forward-model refinement. 
(b)  Schematic visualization of all dihedral angles related to the $^3J$-couplings that are object of the optimization. As an exemplary nucleotide we show Adenosine with all relevant dihedral angles. Since multiple couplings, for which signals are obtainable, can be related using the same Karplus relationship, we list the possible coupled atoms for the respective dihedral angle in the following. $\beta:$ (P-O5$^\prime$-C5$^\prime$-H5$^\prime$, P-O5$^\prime$-C5$^\prime$-H5$^{\prime\prime}$), $\gamma:$   (H5$^\prime$-C5$^\prime$-C4$^\prime$-H4$^\prime$, H5$^{\prime\prime}$-C5$^\prime$-C4$^\prime$-H4$^\prime$), $sugar (\nu_1, \nu_2, \nu_3):$ (H1$^\prime$-C1$^\prime$-C2$^\prime$-H2$^\prime$, H2$^\prime$-C2$^\prime$-C3$^\prime$-H3$^\prime$, H3$^\prime$-C3$^\prime$-C4$^\prime$-H4$^\prime$).
(c) The optimized Karplus parameters are then validated on the UCUCGU hexamer system.}
\end{figure*}

To test the introduced formalism, we report simulation results for a set of RNA tetramers with sequence AAAA, CAAU, CCCC, GACC, UUUU, and for two hexamers with
sequence UCAAUC and UCUCGU (Fig.~\ref{fig:SchematicKarplus}). For the simulation we used the standard OL3 RNA ff~\cite{Cornell1996,Wang2000,Perez2007,Zgarbova2011}  with the van-der-Waals modification of phosphate oxygens developed in Ref.~\onlinecite{Steinbrecher2012} without adjustment of dihedral parameters. As a water model we chose OPC~\cite{IzadiOPC2014}.
This combination has been originally proposed in Ref.~\onlinecite{bergonzo2015improved} and 
already tested on one of the hexamers studied here \cite{bergonzo2022conformational}.
Tetranucleotide simulations were run at salt concentrations corresponding to the experimental conditions to which they are compared to a later point. Therefore, CCCC and GACC were simulated at $\approx$0.09 M KCl, AAAA, UUUU and CAAU at $\approx$0.15 M KCl,  UCAAUC at $\approx$0.11 M KCl, always using the Joung-Cheatham ion parameters~\cite{CheathamIons2008} optimized for TIP4PEwald.
Enhanced sampling simulations of tetranucleotides and tetraloops were run with GROMACS2018~\cite{abraham2015gromacs}. For tetra- and hexanucleotides standard parallel tempering ~\cite{hansmann1997,sugita1999} protocol was used with 24 replicas.
The temperatures were chosen from a geometric distribution ranging from 275 K to 400 K and systems were simulated for 1~$\mu$s per replica.
Only the trajectory closest to 300 K is considered in the following analysis.

\subsection{Experimental data}

All small RNAs simulated in this work can be compared to experimental studies providing NMR data in the form of $^3J$-scalar-couplings as well as observed and unobserved NOEs (uNOEs): AAAA~\cite{condon2015stacking}, CAAU~\cite{condon2015stacking}, CCCC~\cite{tubbs2013nuclear}, GACC~\cite{yildirim2011benchmarking,condon2015stacking}, UUUU~\cite{condon2015stacking}, UCAAUC~\cite{zhao2020nuclear} and UCUCGU~\cite{zhao2022nuclear}.
The choice of simulated RNA systems is based on previous works showing that AAAA, CAAU and CCCC are sequences which possess overpopulated states with intercalated structures, which can hinder correct folding of systems containing them~\cite{mlynsky2020fine}.
The GACC tetramer  was reported to sample intercalated structures not compatible with experiment~\cite{condon2015stacking,bergonzo2015highly} as well,
although this artifact can be significantly decreased using
modified dihedral potentials~\cite{gil2016empirical,Aytenfisu2017,chen2022rna} 
or modified water models~\cite{bergonzo2015improved,Yang2017,bottaro2018conformational,tan2018rna}.
The UUUU system is in disagreement with available experimental NMR data~\cite{condon2015stacking}.
In a study comparing MD simulations with NMR experiments it was found that for CAAU and UCAAUC the simulated dynamics of the termini did not agree with experiment~\cite{zhao2020nuclear}. %

\subsection{Optimized forward models}

We here optimized the forward models used to back-calculate $^3J$ scalar couplings in RNA systems.
We use the functional form proposed in Ref.~\onlinecite{Hecht1963}:
\begin{equation}
^3J(\phi)
= Acos^2(\phi)-Bcos(\phi)+C~.
\label{UniversalKarplusEq}
\end{equation}
where $\phi$ is the involved dihedral angle.
On the basis of previous studies~\cite{bottaro2019barnaba}, we use as an initial guess
the parameters proposed in previous papers: Ref.~\onlinecite{Davies1978} for $\gamma$, Ref.~\onlinecite{Lankhorst1985} for $\beta$,
and Ref.~\onlinecite{condon2015stacking} for the sugar angles.
These parametrizations are only used to set a Gaussian prior on the forward-model parameters $\theta$ for regularization.
The regularization function for each Karplus equation is defined as
\begin{equation}
R = (\frac{A}{2}-\frac{A_0}{2})^2+(B-B_0)^2+(\sqrt{2}(C+\frac{A}{2})-\sqrt{2}(C_0+\frac{A_0}{2}))^2
\label{ExactRegularizationForm}
\end{equation}
where $A_0$, $B_0$, and $C_0$ are the reference parameters.
This is twice the square deviation between the new and the reference Karplus equations averaged over
the angle and thus allows equivalent regularization among the $A$, $B$, and $C$ parameters.
In this work, we consider the functional form of Eq.~\ref{UniversalKarplusEq} using the involved nuclei, and not the corresponding
heavy-atoms, so that no phase shift is required. The exact atoms involved are shown in Fig.~\ref{fig:SchematicKarplus}.

\section{\label{sec:Results}Results}

\subsection{Cross-validation procedure}
\label{Cross_validation_procedure}
\begin{figure*}
\includegraphics[width=0.5\textwidth]{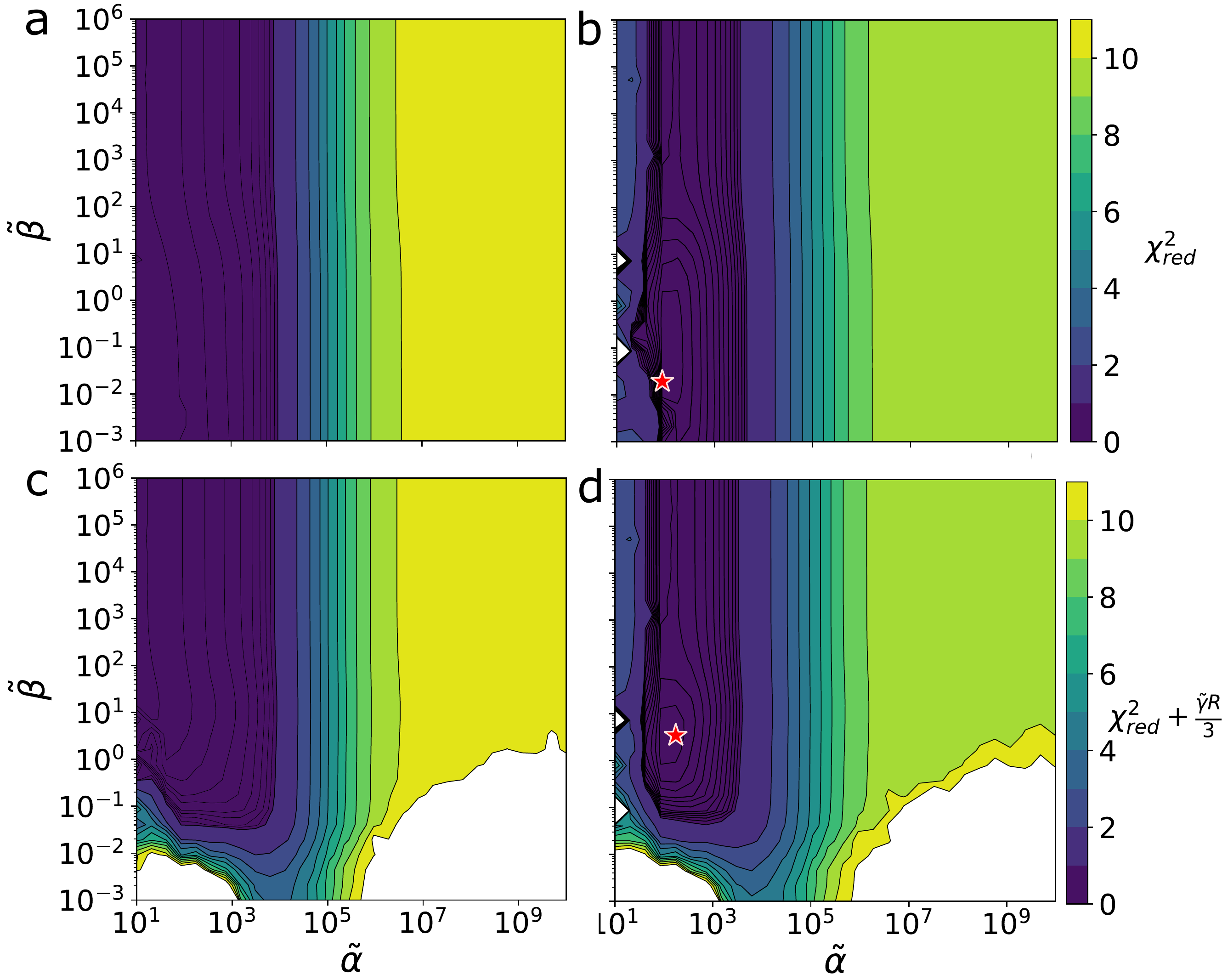}
\caption{\label{fig:CV}
Results of the cross-validation tests using L2 regularization method.
The error function is evaluated on the training (panels a, c) and validation set (panels b, d), using Karplus parameters obtained minimizing the error
and a scan over a wide range of values for both regularization hyperparameters.
The error function for the reference force field is equivalent to the error at maximal hyperparameters. Ranges of both horizontal and vertical axes as well as the sidebar are identical. While panels a and c show the error on the training dataset, panels b, d show the error on the dataset left-out during the training. The definition of the error in a and b is $\chi^2_{\textrm{red}}$ while in c and d it is $\chi^2_{\textrm{red}}+\frac{\tilde{\gamma}R}{3}$ instead, with $\tilde{\gamma}=1\textrm{Hz}^{-2}$,
meaning the average $\chi^2$-error with the added average discrepancy between original and optimized Karplus parameters.}
\end{figure*}

We perform a scan in the space of hyperparameters $\tilde{\alpha}$ and $\tilde{\beta}$ to begin with. In Fig.~\ref{fig:CV}a we collected the average $\chi^2_{\textrm{red}}$ error on the training set. L2 regularization is applied in the outer minimization on the difference between the reference Karplus parameters and any new choice.
L2 regularization is also applied to the inner minimization of $\tilde{\Gamma}$ (Eq.~\ref{eq:gamma-tilde}) as a function of the parameters $\lambda$, as we have shown that this regularization
corresponds to a relative entropy regularization for the cost function.
Therefore, by construction the $\chi^2_{\textrm{red}}$ error is maximal at large hyperparameter choice and equal to the error in the original ensemble. The limit of low hyperparameters
corresponds instead to fitting without any regularization and, accordingly, we expect to observe a decrease in average training $\chi^2_{\textrm{red}}$ when decreasing $\tilde{\alpha}$ and $\tilde{\beta}$ values. 
We notice that the choice of $\tilde{\alpha}$ has an impact on the entire dataset, while changes to $\tilde{\beta}$ impact only the error contribution of $^3J$-couplings. Consequently, the training error is changing more significantly along $\tilde{\alpha}$  than along $\tilde{\beta}$.
When monitoring the average $\chi^2_{\textrm{red}}$ evaluated on the validation set (Fig.~\ref{fig:CV}b), which was left-out during training,
we observe that the cross-validation error is not continuously decreasing with decreasing magnitude in hyperparameters,
indicating that overfitting can occur.
This is especially visible when monitoring the dependence of the cross-validation error on $\tilde{\alpha}$.
The specific values of the hyperparameters $\tilde{\alpha}$ and $\tilde{\beta}$ that minimize the cross-validation error are shown with a star.

The main contribution to the error in Fig.~\ref{fig:CV}a and b is the discrepancy of the simulations with NOE and uNOE experiments.
When using the hyperparameters that minimize the cross-validation error, 
the combined error from those amounts up to $\chi^2_{\textrm{red}}=73.91$, while the error from $^3J$-coupling experiments is only at $\chi^2_{\textrm{red}}=1.54$.
This imbalance might hide possible overfitting issues on the Karplus equations. In addition, and perhaps more importantly,
the overfitting identified here are limited to regions of the dihedral angles that are actually explored in the MD simulation,
and do not consider in any way the agreement between the proposed Karplus parameters and those already used in previous works.
In order to make sure that the trained Karplus parameters do not deviate too much from those proposed in previous works,
we additionally monitor the value of $\chi_{\textrm{red}}^2+\frac{\tilde{\gamma}R}{3}$, where $R$ is the regularization penalty for the Karplus parameters
which is here divided by three to consider that we are modifying three Karplus equations. The prefactor $\tilde{\gamma}$ is an additional
hyperparameter that takes into account the relative importance of finding Karplus equations that agree with the currently analyzed experiments
and Karplus equations that agree with the literature.
Heuristically, we found that $\tilde{\gamma}=1 Hz^{-2}$ allows to identify Karplus parameters that do not deviate too much from those in the literature.
This value corresponds to (a) assigning an uncertainty of $\approx$ 0.7 Hz to the equations reported in the literature and (b)
assigning the same relative weight to the cross validation against left-out datapoints and to the cross validation against the reference Karplus parameters.
In Fig.~\ref{fig:CV}c we observe a decrease in average training $\chi_{\textrm{red}}^2+\frac{\tilde{\gamma}R}{3}$ when decreasing $\tilde{\alpha}$ and $\tilde{\beta}$ values. However, in the limit of small $\tilde{\alpha}$ and $\tilde{\beta}$ the behaviour changes and a strong increase in the error can be seen. This indicates a significant change in the Karplus parameters.
In Fig.~\ref{fig:CV}d we see an accentuated trend of what was already seen in Fig.~\ref{fig:CV}b. In the limit of small $\tilde{\beta}$ one would identify new Karplus parameters, which would be non-transferable to new data points and, additionally, would deviate significantly from those reported in the literature.

\subsection{Discovery of optimized Karplus parameters}
\label{Discovery_opt_Karplus_params}
\begin{figure*}
\includegraphics[width=\textwidth]{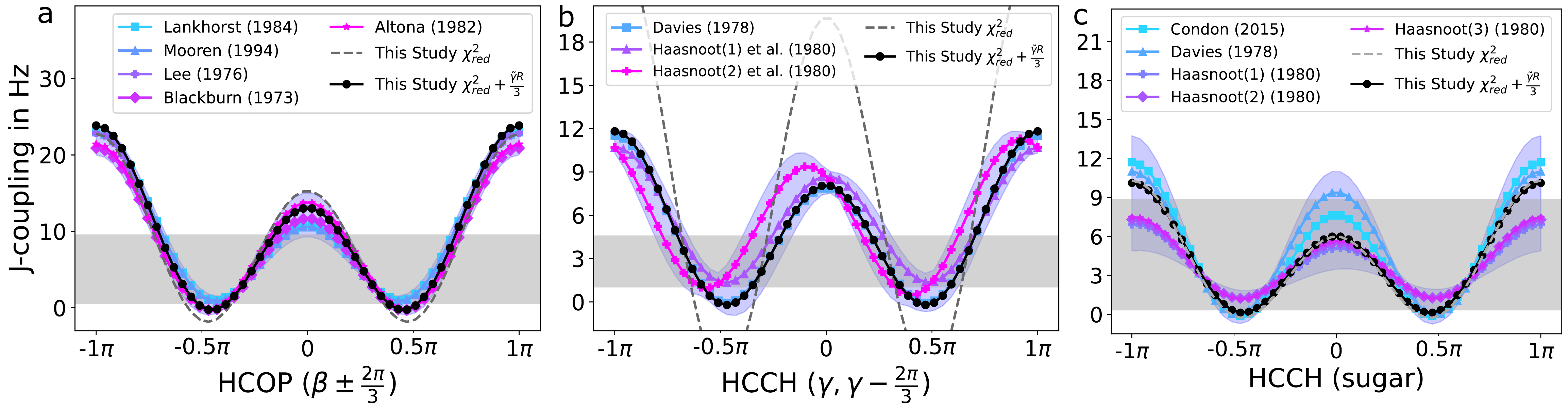}
\caption{\label{fig:OptimizedKarplus}
Comparison of the optimized Karplus parameters identified in this study to Karplus curves corresponding to parameter sets previously proposed in the literature including the ones used as starting parameters for the fitting ($\beta$: Lankhorst~\cite{Lankhorst1985}-Mooren~\cite{mooren1994solution}-Lee~\cite{lee1976aqueous}-Blackburn~\cite{blackburn1973proton}-Altona~\cite{altona1982conformational}, $\gamma$: Davies~\cite{Davies1978}-Haasnoot~\cite{condon2015stacking}, $sugar$: Condon~\cite{condon2015stacking}-Davies~\cite{Davies1978}-Haasnoot~\cite{condon2015stacking}). Light-grey horizontal bar corresponds to minimal and maximal experimental values for the respective $^3J$-coupling considering the entire training database. In order to show overfitting issues, the Karplus curves resulting from optimization with hyperparameters chosen based on Fig.~\ref{fig:CV}b are also included.}
\end{figure*}

The minimum cross-validation error identified in Fig.~\ref{fig:CV}d, as shown in the previous section, is the criterion to choose the strength of a specific form of the regularization term for inner and outer minimization.
In Table~\ref{tab_KarplusParameters} and Fig.~\ref{fig:OptimizedKarplus} we compare the Karplus parameters and their resulting Karplus curves before and after the optimization procedure is completed. From Table~\ref{tab_KarplusParameters} we can appreciate significant differences in parameter values for the sets related to the $\beta$ and the $sugar$ dihedral angles
(root-mean-square-deviation is $\approx$ 1 Hz).
Optimization of the Karplus parameters for $\gamma$ leads to small changes instead
(root-mean-square-deviation is $\approx$ 0.15 Hz).
Fig.~\ref{fig:OptimizedKarplus} additionally reports the curves obtained with other sets of parameters reported in the literature
as well as the range of experimental value spanned by the datapoints used in this work.
Optimized Karplus equations are shown as obtained using both options for the definition of the cross-validation error,
namely $\chi^2_{\textrm{red}}$, reporting the disagreement with cross-validation datapoints, and $\chi^2_{\textrm{red}}+\frac{\tilde{\gamma}R}{3}$, which also reports on the disagreement
with Karplus equations taken from the literature.
Importantly, a significant difference can be spotted for the Karplus curves optimized based on the 2 different definitions. While the ones obtained based on the $\chi^2$-error suffer from overfitting issues, the parameter set obtained with the $\chi^2_{\textrm{red}}+\frac{\tilde{\gamma}R}{3}$ is closer to the experimental range considered in this study and, as expected, lies within the standard deviation of Karplus curves previously published.
The departure from previously determined Karplus curves is particularly evident for the $\gamma$ Karplus parameters,
whereas it is very mild for  $sugar$ parameters. This can be explained by the fact that the distribution of the
$\gamma$ angle is peaked (compare SI Fig.~1 for histograms of the dihedral angles cumulated over all systems) %
with the consequence that the cross-validation test is not sensitive to modification of the Karplus curves in regions with
very low population.
The sampling of the dihedral angle used with $sugar$ related parameters spans instead entire range between $-\pi$ and $\pi$ reducing overfitting on local regions in dihedral angle space.

 The results of optimizing Karplus parameters and performing maximum entropy reweighting at the optimal hyperparameter identified via cross-validation and using the entire database are reported in Table~\ref{tab_KarplusTrainingErrors}.
The errors in the original ensemble are very large for NOEs and are significantly reduced by the reweighting procedure.
The errors are moderate for $^3J$ scalar couplings, and are further reduced when using the optimized Karplus curves.
Notably, the use of the optimized Karplus equations further reduces the discrepancy between simulation and experiment.

\subsection{Validation of the optimized Karplus parameters on left-out data}

The weights computed using the procedure above were also used to evaluate the discrepancy between simulation and experiment
in a set of data points, 20$\%$ of the entire dataset, which was completely left-out from the optimization procedures.
This validation set allows testing the fitting results on new unknown data for the same systems considered during optimizations.
The original $\chi^2_{\textrm{red}}$ containing NOEs and uNOEs is 10.16. Using the weights obtained during maximum entropy reweighting at $\tilde{\beta}=\infty$ decreases the error to 1.94. The evaluation of the error after reweighting with the optimized Karplus parameters obtained at  optimal hyperparameters leads to the same contribution of NOE and uNOEs, however the contribution coming from $^3J$-couplings is further reduced.
As can be seen from SI Fig.~2, this is due to the fact that the weights are not changing significantly when reweighting the ensemble with the optimized Karplus parameters.
For the specific systems, this is a consequence of the strong dominance of the non-refined NOEs and uNOEs data over the refined scalar couplings.

\begin{table}%
\caption{In this table we collect the original ($\beta$~\cite{Lankhorst1985}, $\gamma$~\cite{Davies1978} and  $sugar$~\cite{condon2015stacking} ) and the optimized set of parameters in this work. The functional form relating observed $^3J$ vicinal proton coupling to the dihedral angle is: $^3J(HH)=Acos^2(\phi)-Bcos(\phi)+C$. 
Coefficients are reported in Hz.
}
\makebox[0.5\textwidth]{
\def\arraystretch{1.3}
\begin{tabular}{ c|c|c|c|c   }
\textbf{Name} & \textbf{A} & \textbf{B} & \textbf{C}& \textbf{Ref.}\\
\hline 
\multicolumn{5}{c}{\textit{$\beta \pm \frac{2\pi}{3}$}} \\
\hline 
original  & 15.3 & -6.1 & 1.6 & Lankhorst~\cite{Lankhorst1985}, 1985\\
\hline 
optimized  & 18.34 & -5.39 & 0.11 & this study \\
\hline 
\multicolumn{5}{c}{\textit{$\gamma,\gamma-\frac{2\pi}{3}$}} \\
\hline 
original  & 9.7 & -1.8 & 0 & Davies~\cite{Davies1978}, 1978 \\
\hline 
optimized  & 10.07 & -1.87 & -0.13 & this study \\
\hline
\multicolumn{5}{c}{\textit{$sugar$}} \\
\hline 
original  & 9.67 & -2.03 & 0 & Condon~\cite{condon2015stacking}, 2015 \\
\hline 
optimized  & 7.81 & -2.05 & 0.25 & this study \\
\end{tabular}}
\label{tab_KarplusParameters}
\end{table}

\begin{table}%
\caption{
Average $\chi^2_{\textrm{red}}$-errors and Kish size evaluated on the training database as well as on the left-out data used for validation. We report both the direct results of the original simulations and the results predicted by reweighting those simulations at different choices of the hyperparameters. Reweighting is performed at the hyperparameters $\tilde{\alpha}$ and $\tilde{\beta}$ specified in the first 2 rows of the table. The optimized Karplus parameters obtained with hyperparameters chosen based on Fig.~\ref{fig:CV}d reduce the average $\chi^2_{\textrm{red}}$-errors in training and validation dataset the most.
}
\makebox[0.5\textwidth]{
\def\arraystretch{1.3}
\begin{tabular}{ c|c|c|c   }
 & \textbf{orig. ensemble} & \textbf{orig. Karplus} & \textbf{opt. Karplus}\\
\hline 
$\tilde{\alpha}$  & $\infty$ &  174 &  174\\
\hline 
$\tilde{\beta}$  & $\infty$ &  $\infty$ &  3.4  \\
\hline 
\multicolumn{4}{c}{\textit{Training dataset}} \\
\hline 
$\chi^2_{\textrm{red}}$ $^3J$-couplings only   & 1.54 &  0.82 &  0.53\\
\hline
$\chi^2_{\textrm{red}}$ NOEs only   & 73.91 &  0.22 &  0.22\\
\hline 
\multicolumn{4}{c}{\textit{Validation dataset}} \\
\hline 
$\chi^2_{\textrm{red}}$ $^3J$-couplings only   & 9.66 &  4.22 &  3.11\\
\hline 
$\chi^2_{\textrm{red}}$ NOEs only   & 10.16 &  1.94 &  1.94\\
\end{tabular}}
\label{tab_KarplusTrainingErrors}
\end{table}

\subsection{Validation of the optimized Karplus parameters on a separate system}
\label{Validation_opt_Karplus_params}
\begin{figure*}
\includegraphics[width=0.5\textwidth]{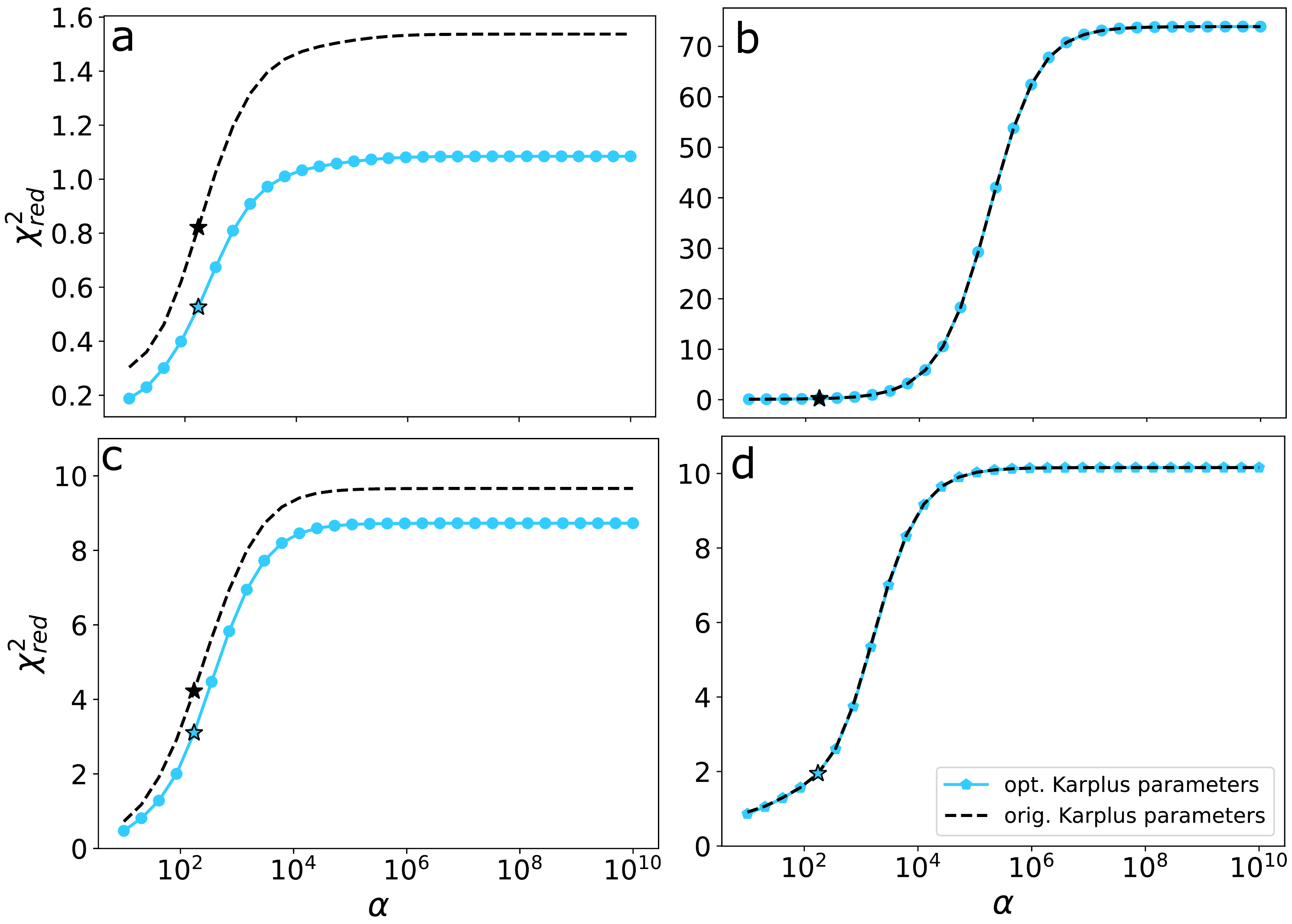}
\caption{\label{fig:Validation}
Sensitivity test evaluating the average $\chi^2$ contributions with the original parameters and the optimal parameters found in this study, considering $^3J$-couplings (a, c), NOEs and uNOEs (b, d) for the training systems (a, b) as well as the new independent UCUCGU hexamer system (c, d). The ensembles are reweighted via maximum entropy method over a range of L2 regularization strengths $\tilde{\alpha}$. The $\tilde{\alpha}$ at which the new Karplus parameters are obtained is indicated by stars. 'orig.' and 'opt.' denote which Karplus parameters are used during reweighting. 
The optimized Karplus parameters are performing well on the training dataset and more importantly are transferable to the dataset completely unknown during the optimization procedure. As intended, the NOE and uNOE errors are not influenced by the optimization of the Karplus parameters.}
\end{figure*}

The optimized set of Karplus parameters seems in agreement with the available experimental data and do not depart significantly from sets previously reported. Therefore, we proceed by evaluating their performance on a new system not considered during training, namely the UCUCGU hexamer. For this RNA system experimental NMR data, consisting of $^3J$-coupling, NOEs and uNOEs, are made available in Ref.~\onlinecite{zhao2022nuclear} such that comparisons to simulations can be made. 
Instead of performing cross-validation to identify the optimal hyperparameter $\tilde{\alpha}$ for maximum entropy reweighting we perform a scan over a range of regularization strengths. The results of this analysis are collected in Fig.~\ref{fig:Validation} and suggest that this set of parameters allows better agreement of training and validation data to their corresponding experimental data when compared to our choice of original Karplus parameters. Not surprisingly, the new Karplus parameters optimized on the training database reduce the $\chi^2$ contribution of $^3J$-couplings significantly over the entire range of regularization strengths. From Fig.~\ref{fig:Validation}a we notice that NOEs and uNOEs are the main contribution to the $\chi^2$ in the limit of large $\tilde{\alpha}$, while for smaller regularization strength $^3J$-couplings exhibit equal contribution to $\chi^2$. The set of Karplus parameters optimized on a $\chi^2$ with these relative error contributions do not change the increase in agreement which can be achieved for NOEs and uNOEs for the training data by performing maximum entropy reweighting, while by construction increasing the agreement with experimental $^3J$-coupling data (compare Fig.~\ref{fig:Validation}a and b). Focusing on the independent UCUCGU system in Fig.~\ref{fig:Validation}c and d, and considering the contributions of NOEs, uNOEs and $^3J$-couplings to the $\chi^2$ one can see that in this specific system the disagreement between simulation and experiments is significantly larger for the $^3J$-couplings compared to the NOEs and uNOEs. However, applying the optimized Karplus parameter set is showing similar results as for the training database by reducing the  $\chi^2$ in $^3J$-couplings significantly while simultaneously allowing the same increase in agreement with experimental NOE data as the original Karplus parameters.

\section{\label{sec:Discussion}Discussion}

In this work, we combine the BioEn method, introduced in Ref.~\onlinecite{hummer2015bayesian}, with the optimization of the empirical Karplus parameters for $\beta$ (H-C-O-P), $\gamma$ (H-C-C-H) and $sugar$ (H-C-C-H). These forward models have been introduced in~\cite{Lankhorst1985,Davies1978,condon2015stacking} and were used in previous works to improve the agreement of simulations with experimental data~\cite{cesari2016combining,bottaro2018conformational,cesari2019fitting,bottaro2020integrating,frohlking2022automatic}.
Specifically, we combine MD simulations with experimental data for a database of the RNA systems AAAA, CAAU, CCCC, GACC, UUU and UCAAUC to fit the Karplus parameters while simultaneously reweighting the simulated ensemble using the maximum entropy method. The optimized Karplus parameters of this study allow for a significant increase in agreement between simulations and experiments for all RNA systems and are transferable to an additional system (UCUCGU) not seen during training.
We apply a rigorous regularization protocol, which probes for errors due to finite data space in multiple directions. We test the L2 regularization strategy, which is standard in the machine learning community and can be directly interpreted as a Gaussian prior distribution on the parameters aimed at keeping them small and close to their \textit{a priori} parametrization.
Whereas the importance of dynamics in the determination of Karplus parameters has been recognized since a long time for protein systems
\cite{lindorff2005interpreting}, we are not aware of similar attempts done for nucleic acids. Indeed, the most commonly used
parametrizations, that we here used as a starting point, are all based on the analysis of static structures.

A weakness of the current reweighting-based approach is that it might statistically inefficient \cite{shen2008statistical}.
The investigated oligomers are sufficiently simple and with relatively accurate initial ensembles so that this
is not creating significant issues. For more complex systems, methods based on on-the-fly restraining might be more effective \cite{rangan2018determination}.
Interestingly, in the metainference approach \cite{bonomi2016metainference}, (systematic) forward-model errors
are treated on par with (random) experimental errors. Whereas one could imagine to adjust forward models
on the fly, it is not clear how to do it efficiently without the need to perform the simulation of all
the studied systems simultaneously and with an explicit coupling, as done for instance in Ref.~\onlinecite{cesari2016combining}. The reweighting approach used here allows
to combine multiple systems a posteriori, and so is less limited in this respect.

From a mathematical point of view,
the optimizations performed in this work correspond to solving a minimax problem (Eq.~\ref{eq:minimax}). Given the relevance of minimax problems in current machine learning
literature (see, e.g., \cite{sanjabi2018convergence}), it might be interesting to test more advanced minimax algorithms, especially if a larger number of forward-model parameters
are to be simultaneously optimized. The nested minimization algorithm used here was sufficient for the current application.

The interpretation of the optimized Karplus parameters is straightforward, as they directly transform a dihedral angle between a specific choice of atoms into a corresponding $^3J$-coupling signal. Ideally, the sampled dihedral angle configurations span the entire range between $-\pi$ and $\pi$. In this study, however, the $\gamma$ angle is missing data points in some regions of the dihedral space (compare SI Fig.~1). In those cases, overfitting parameters on the regions with data points can occur, meaning we are introducing extrapolation errors in the regions in which data points are missing. Monitoring simultaneously disagreement of the validation set with experiments and how much the optimized Karplus parameters depart from the initial choice during cross-validation allows us to avoid such overfitting due to finite sampling or data points. We herein split the data into 70$\%$ for training and 30$\%$ for validation.
The splitting is done so as to combine cross-validation on the trajectory and cross-validation on observables.
Performed in this way, cross-validation simultaneously tests how well the fitted corrections are transferable to newly sampled data points along the trajectory and new experimental measurements. We identify overfitting issues in absence of regularization for both Karplus parameters and the ensemble reweighting, emphasizing the importance of
using a regularization term.

The idea of tuning free parameters related to forward models can be extended to other experimental observables, where parameters could
be either transferable or non-transferable between different experiments.
We here focused on transferable Karplus parameters for $^3J$-coupling experiments.
However, it is instructive to consider how cross-peaks are used to generate upper (or lower) distance bounds for observed (or unobserved) NOEs.
Specifically, peak intensities are calibrated iteratively, and different protons are assigned different calibration constants
\cite{herrmann2002protein,lange2014automatic},
 for which more optimal choices could exist. The method introduced here could be used to calibrate them in a fashion consistent with MD ensembles during the ensemble refinement step.

A similar issue appears in SAXS data, where a scaling factor and a shift might be necessary to maximize the agreement with experiment and
compensate for imperfect background subtraction \cite{pesce2021refining}. Pesce and Lindorff-Larsen proposed
to determine these coefficients simultaneously with the ensemble refinement step, in an approach that is similar to the one introduce here.
The main differences are that (a) in Ref.~\onlinecite{pesce2021refining} the forward models and the ensembles are refined alternatively, rather than concurrently,
and that (b) the optimization is done on the $\chi^2$ rather than on the BioEN cost function.
Another important novelty introduced here is the application of a regularization term on the force-field parameters,
which facilitates obtaining parameters that are then transferable to other systems.
We notice that, in a previous work, we were adapting the maximum-entropy formalism so as to be able to
enforce SAXS data without specifying the scaling factor \cite{bernetti2021reweighting}.
This however made it difficult to add a suitable regularization term.
The approach of Ref.~\onlinecite{pesce2021refining}, or the approach introduced here, would be a better way to tackle the same issue.

Future works will investigate the possibility to add force field fitting~\cite{cesari2019fitting,frohlking2022automatic} to our combination of forward model optimization and ensemble refinement. This would translate into a framework which can integrate molecular simulations and experimental data considering that experiments, forward models and force fields might have errors.

\section{Supporting Information}
At \url{https://github.com/bussilab/forward-model-optimization} we provided the SI figures 1 and 2 as 'SI\_Figures\_1\_2.ipynb' as well as all scripts that are allowing one to obtain the results in this study.

\begin{acknowledgments}
We acknowledge J{\"u}rgen K{\"o}finger and Kenno Vanommeslaeghe
for reading the PhD thesis of T.F., including a preliminary version of this work, and providing several useful suggestions. We also acknowledge Gerhard Hummer for useful discussion.
\end{acknowledgments}

\section*{Data availability}

The data that support the findings of this study are openly available
at \url{https://zenodo.org/record/7746293}.
Analysis scripts
can be found at \url{https://github.com/bussilab/forward-model-optimization}.

\bibliography{main}

\begin{thebibliography}{75}%
\makeatletter
\providecommand \@ifxundefined [1]{%
 \@ifx{#1\undefined}
}%
\providecommand \@ifnum [1]{%
 \ifnum #1\expandafter \@firstoftwo
 \else \expandafter \@secondoftwo
 \fi
}%
\providecommand \@ifx [1]{%
 \ifx #1\expandafter \@firstoftwo
 \else \expandafter \@secondoftwo
 \fi
}%
\providecommand \natexlab [1]{#1}%
\providecommand \enquote  [1]{``#1''}%
\providecommand \bibnamefont  [1]{#1}%
\providecommand \bibfnamefont [1]{#1}%
\providecommand \citenamefont [1]{#1}%
\providecommand \href@noop [0]{\@secondoftwo}%
\providecommand \href [0]{\begingroup \@sanitize@url \@href}%
\providecommand \@href[1]{\@@startlink{#1}\@@href}%
\providecommand \@@href[1]{\endgroup#1\@@endlink}%
\providecommand \@sanitize@url [0]{\catcode `\\12\catcode `\$12\catcode
  `\&12\catcode `\#12\catcode `\^12\catcode `\_12\catcode `\%12\relax}%
\providecommand \@@startlink[1]{}%
\providecommand \@@endlink[0]{}%
\providecommand \url  [0]{\begingroup\@sanitize@url \@url }%
\providecommand \@url [1]{\endgroup\@href {#1}{\urlprefix }}%
\providecommand \urlprefix  [0]{URL }%
\providecommand \Eprint [0]{\href }%
\providecommand \doibase [0]{http://dx.doi.org/}%
\providecommand \selectlanguage [0]{\@gobble}%
\providecommand \bibinfo  [0]{\@secondoftwo}%
\providecommand \bibfield  [0]{\@secondoftwo}%
\providecommand \translation [1]{[#1]}%
\providecommand \BibitemOpen [0]{}%
\providecommand \bibitemStop [0]{}%
\providecommand \bibitemNoStop [0]{.\EOS\space}%
\providecommand \EOS [0]{\spacefactor3000\relax}%
\providecommand \BibitemShut  [1]{\csname bibitem#1\endcsname}%
\let\auto@bib@innerbib\@empty
%</preamble>
\bibitem [{\citenamefont {Hollingsworth}\ and\ \citenamefont
  {Dror}(2018)}]{hollingsworth2018molecular}%
  \BibitemOpen
  \bibfield  {author} {\bibinfo {author} {\bibfnamefont {S.~A.}\ \bibnamefont
  {Hollingsworth}}\ and\ \bibinfo {author} {\bibfnamefont {R.~O.}\ \bibnamefont
  {Dror}},\ }\bibfield  {title} {\enquote {\bibinfo {title} {Molecular dynamics
  simulation for all},}\ }\href@noop {} {\bibfield  {journal} {\bibinfo
  {journal} {Neuron}\ }\textbf {\bibinfo {volume} {99}},\ \bibinfo {pages}
  {1129--1143} (\bibinfo {year} {2018})}\BibitemShut {NoStop}%
\bibitem [{\citenamefont {Bonomi}\ \emph {et~al.}(2017)\citenamefont {Bonomi},
  \citenamefont {Heller}, \citenamefont {Camilloni},\ and\ \citenamefont
  {Vendruscolo}}]{bonomi2017principles}%
  \BibitemOpen
  \bibfield  {author} {\bibinfo {author} {\bibfnamefont {M.}~\bibnamefont
  {Bonomi}}, \bibinfo {author} {\bibfnamefont {G.~T.}\ \bibnamefont {Heller}},
  \bibinfo {author} {\bibfnamefont {C.}~\bibnamefont {Camilloni}}, \ and\
  \bibinfo {author} {\bibfnamefont {M.}~\bibnamefont {Vendruscolo}},\
  }\bibfield  {title} {\enquote {\bibinfo {title} {Principles of protein
  structural ensemble determination},}\ }\href@noop {} {\bibfield  {journal}
  {\bibinfo  {journal} {Curr. Opin. Struct. Biol.}\ }\textbf {\bibinfo {volume}
  {42}},\ \bibinfo {pages} {106--116} (\bibinfo {year} {2017})}\BibitemShut
  {NoStop}%
\bibitem [{\citenamefont {Bottaro}\ and\ \citenamefont
  {Lindorff-Larsen}(2018)}]{bottaro2018biophysical}%
  \BibitemOpen
  \bibfield  {author} {\bibinfo {author} {\bibfnamefont {S.}~\bibnamefont
  {Bottaro}}\ and\ \bibinfo {author} {\bibfnamefont {K.}~\bibnamefont
  {Lindorff-Larsen}},\ }\bibfield  {title} {\enquote {\bibinfo {title}
  {Biophysical experiments and biomolecular simulations: A perfect match?}}\
  }\href@noop {} {\bibfield  {journal} {\bibinfo  {journal} {Science}\ }\textbf
  {\bibinfo {volume} {361}},\ \bibinfo {pages} {355--360} (\bibinfo {year}
  {2018})}\BibitemShut {NoStop}%
\bibitem [{\citenamefont {Bernetti}\ and\ \citenamefont
  {Bussi}(2023)}]{bernetti2023integrating}%
  \BibitemOpen
  \bibfield  {author} {\bibinfo {author} {\bibfnamefont {M.}~\bibnamefont
  {Bernetti}}\ and\ \bibinfo {author} {\bibfnamefont {G.}~\bibnamefont
  {Bussi}},\ }\bibfield  {title} {\enquote {\bibinfo {title} {Integrating
  experimental data with molecular simulations to investigate {RNA} structural
  dynamics},}\ }\href@noop {} {\bibfield  {journal} {\bibinfo  {journal} {Curr.
  Opin. Struct. Biol.}\ }\textbf {\bibinfo {volume} {78}},\ \bibinfo {pages}
  {102503} (\bibinfo {year} {2023})}\BibitemShut {NoStop}%
\bibitem [{\citenamefont {Norgaard}, \citenamefont {Ferkinghoff-Borg},\ and\
  \citenamefont {Lindorff-Larsen}(2008)}]{norgaard2008experimental}%
  \BibitemOpen
  \bibfield  {author} {\bibinfo {author} {\bibfnamefont {A.~B.}\ \bibnamefont
  {Norgaard}}, \bibinfo {author} {\bibfnamefont {J.}~\bibnamefont
  {Ferkinghoff-Borg}}, \ and\ \bibinfo {author} {\bibfnamefont
  {K.}~\bibnamefont {Lindorff-Larsen}},\ }\bibfield  {title} {\enquote
  {\bibinfo {title} {Experimental parameterization of an energy function for
  the simulation of unfolded proteins},}\ }\href@noop {} {\bibfield  {journal}
  {\bibinfo  {journal} {Biophys. J.}\ }\textbf {\bibinfo {volume} {94}},\
  \bibinfo {pages} {182--192} (\bibinfo {year} {2008})}\BibitemShut {NoStop}%
\bibitem [{\citenamefont {Li}\ and\ \citenamefont
  {Br{\"u}schweiler}(2011)}]{li2011iterative}%
  \BibitemOpen
  \bibfield  {author} {\bibinfo {author} {\bibfnamefont {D.-W.}\ \bibnamefont
  {Li}}\ and\ \bibinfo {author} {\bibfnamefont {R.}~\bibnamefont
  {Br{\"u}schweiler}},\ }\bibfield  {title} {\enquote {\bibinfo {title}
  {Iterative optimization of molecular mechanics force fields from {NMR} data
  of full-length proteins},}\ }\href@noop {} {\bibfield  {journal} {\bibinfo
  {journal} {J. Chem. Theory Comput.}\ }\textbf {\bibinfo {volume} {7}},\
  \bibinfo {pages} {1773--1782} (\bibinfo {year} {2011})}\BibitemShut {NoStop}%
\bibitem [{\citenamefont {Wang}, \citenamefont {Chen},\ and\ \citenamefont
  {Van~Voorhis}(2012)}]{wang2012systematic}%
  \BibitemOpen
  \bibfield  {author} {\bibinfo {author} {\bibfnamefont {L.-P.}\ \bibnamefont
  {Wang}}, \bibinfo {author} {\bibfnamefont {J.}~\bibnamefont {Chen}}, \ and\
  \bibinfo {author} {\bibfnamefont {T.}~\bibnamefont {Van~Voorhis}},\
  }\bibfield  {title} {\enquote {\bibinfo {title} {Systematic parametrization
  of polarizable force fields from quantum chemistry data},}\ }\href@noop {}
  {\bibfield  {journal} {\bibinfo  {journal} {J. Chem. Theory Comput.}\
  }\textbf {\bibinfo {volume} {9}},\ \bibinfo {pages} {452--460} (\bibinfo
  {year} {2012})}\BibitemShut {NoStop}%
\bibitem [{\citenamefont {Wang}, \citenamefont {Martinez},\ and\ \citenamefont
  {Pande}(2014)}]{wang2014building}%
  \BibitemOpen
  \bibfield  {author} {\bibinfo {author} {\bibfnamefont {L.-P.}\ \bibnamefont
  {Wang}}, \bibinfo {author} {\bibfnamefont {T.~J.}\ \bibnamefont {Martinez}},
  \ and\ \bibinfo {author} {\bibfnamefont {V.~S.}\ \bibnamefont {Pande}},\
  }\bibfield  {title} {\enquote {\bibinfo {title} {Building force fields: an
  automatic, systematic, and reproducible approach},}\ }\href@noop {}
  {\bibfield  {journal} {\bibinfo  {journal} {J. Phys. Chem. Lett.}\ }\textbf
  {\bibinfo {volume} {5}},\ \bibinfo {pages} {1885--1891} (\bibinfo {year}
  {2014})}\BibitemShut {NoStop}%
\bibitem [{\citenamefont {Cesari}, \citenamefont {Gil-Ley},\ and\ \citenamefont
  {Bussi}(2016)}]{cesari2016combining}%
  \BibitemOpen
  \bibfield  {author} {\bibinfo {author} {\bibfnamefont {A.}~\bibnamefont
  {Cesari}}, \bibinfo {author} {\bibfnamefont {A.}~\bibnamefont {Gil-Ley}}, \
  and\ \bibinfo {author} {\bibfnamefont {G.}~\bibnamefont {Bussi}},\ }\bibfield
   {title} {\enquote {\bibinfo {title} {Combining simulations and solution
  experiments as a paradigm for {RNA} force field refinement},}\ }\href@noop {}
  {\bibfield  {journal} {\bibinfo  {journal} {J. Chem. Theory Comput.}\
  }\textbf {\bibinfo {volume} {12}},\ \bibinfo {pages} {6192--6200} (\bibinfo
  {year} {2016})}\BibitemShut {NoStop}%
\bibitem [{\citenamefont {Cesari}\ \emph {et~al.}(2019)\citenamefont {Cesari},
  \citenamefont {Bottaro}, \citenamefont {Lindorff-Larsen}, \citenamefont
  {Ban{\'a}{\v{s}}}, \citenamefont {{\v{S}}poner},\ and\ \citenamefont
  {Bussi}}]{cesari2019fitting}%
  \BibitemOpen
  \bibfield  {author} {\bibinfo {author} {\bibfnamefont {A.}~\bibnamefont
  {Cesari}}, \bibinfo {author} {\bibfnamefont {S.}~\bibnamefont {Bottaro}},
  \bibinfo {author} {\bibfnamefont {K.}~\bibnamefont {Lindorff-Larsen}},
  \bibinfo {author} {\bibfnamefont {P.}~\bibnamefont {Ban{\'a}{\v{s}}}},
  \bibinfo {author} {\bibfnamefont {J.}~\bibnamefont {{\v{S}}poner}}, \ and\
  \bibinfo {author} {\bibfnamefont {G.}~\bibnamefont {Bussi}},\ }\bibfield
  {title} {\enquote {\bibinfo {title} {Fitting corrections to an {RNA} force
  field using experimental data},}\ }\href@noop {} {\bibfield  {journal}
  {\bibinfo  {journal} {J. Chem. Theory Comput.}\ }\textbf {\bibinfo {volume}
  {15}},\ \bibinfo {pages} {3425--3431} (\bibinfo {year} {2019})}\BibitemShut
  {NoStop}%
\bibitem [{\citenamefont {Fr{\"o}hlking}\ \emph {et~al.}(2020)\citenamefont
  {Fr{\"o}hlking}, \citenamefont {Bernetti}, \citenamefont {Calonaci},\ and\
  \citenamefont {Bussi}}]{frohlking2020toward}%
  \BibitemOpen
  \bibfield  {author} {\bibinfo {author} {\bibfnamefont {T.}~\bibnamefont
  {Fr{\"o}hlking}}, \bibinfo {author} {\bibfnamefont {M.}~\bibnamefont
  {Bernetti}}, \bibinfo {author} {\bibfnamefont {N.}~\bibnamefont {Calonaci}},
  \ and\ \bibinfo {author} {\bibfnamefont {G.}~\bibnamefont {Bussi}},\
  }\bibfield  {title} {\enquote {\bibinfo {title} {Toward empirical force
  fields that match experimental observables},}\ }\href@noop {} {\bibfield
  {journal} {\bibinfo  {journal} {J. Chem. Phys.}\ }\textbf {\bibinfo {volume}
  {152}},\ \bibinfo {pages} {230902} (\bibinfo {year} {2020})}\BibitemShut
  {NoStop}%
\bibitem [{\citenamefont {K{\"o}finger}\ and\ \citenamefont
  {Hummer}(2021)}]{kofinger2021empirical}%
  \BibitemOpen
  \bibfield  {author} {\bibinfo {author} {\bibfnamefont {J.}~\bibnamefont
  {K{\"o}finger}}\ and\ \bibinfo {author} {\bibfnamefont {G.}~\bibnamefont
  {Hummer}},\ }\bibfield  {title} {\enquote {\bibinfo {title} {Empirical
  optimization of molecular simulation force fields by {B}ayesian inference},}\
  }\href@noop {} {\bibfield  {journal} {\bibinfo  {journal} {Eur. Phys. J. B}\
  }\textbf {\bibinfo {volume} {94}},\ \bibinfo {pages} {245} (\bibinfo {year}
  {2021})}\BibitemShut {NoStop}%
\bibitem [{\citenamefont {Fr{\"o}hlking}\ \emph {et~al.}(2022)\citenamefont
  {Fr{\"o}hlking}, \citenamefont {Ml{\`y}nsk{\`y}}, \citenamefont
  {Jane{\v{c}}ek}, \citenamefont {K{\"u}hrov{\'a}}, \citenamefont {Krepl},
  \citenamefont {Ban{\'a}{\v{s}}}, \citenamefont {{\v{S}}poner},\ and\
  \citenamefont {Bussi}}]{frohlking2022automatic}%
  \BibitemOpen
  \bibfield  {author} {\bibinfo {author} {\bibfnamefont {T.}~\bibnamefont
  {Fr{\"o}hlking}}, \bibinfo {author} {\bibfnamefont {V.}~\bibnamefont
  {Ml{\`y}nsk{\`y}}}, \bibinfo {author} {\bibfnamefont {M.}~\bibnamefont
  {Jane{\v{c}}ek}}, \bibinfo {author} {\bibfnamefont {P.}~\bibnamefont
  {K{\"u}hrov{\'a}}}, \bibinfo {author} {\bibfnamefont {M.}~\bibnamefont
  {Krepl}}, \bibinfo {author} {\bibfnamefont {P.}~\bibnamefont
  {Ban{\'a}{\v{s}}}}, \bibinfo {author} {\bibfnamefont {J.}~\bibnamefont
  {{\v{S}}poner}}, \ and\ \bibinfo {author} {\bibfnamefont {G.}~\bibnamefont
  {Bussi}},\ }\bibfield  {title} {\enquote {\bibinfo {title} {Automatic
  learning of hydrogen-bond fixes in an {AMBER} {RNA} force field},}\
  }\href@noop {} {\bibfield  {journal} {\bibinfo  {journal} {J. Chem. Theory
  Comput.}\ }\textbf {\bibinfo {volume} {18}},\ \bibinfo {pages} {4490--4502}
  (\bibinfo {year} {2022})}\BibitemShut {NoStop}%
\bibitem [{\citenamefont {Cavalli}, \citenamefont {Camilloni},\ and\
  \citenamefont {Vendruscolo}(2013)}]{cavalli2013molecular}%
  \BibitemOpen
  \bibfield  {author} {\bibinfo {author} {\bibfnamefont {A.}~\bibnamefont
  {Cavalli}}, \bibinfo {author} {\bibfnamefont {C.}~\bibnamefont {Camilloni}},
  \ and\ \bibinfo {author} {\bibfnamefont {M.}~\bibnamefont {Vendruscolo}},\
  }\bibfield  {title} {\enquote {\bibinfo {title} {Molecular dynamics
  simulations with replica-averaged structural restraints generate structural
  ensembles according to the maximum entropy principle},}\ }\href@noop {}
  {\bibfield  {journal} {\bibinfo  {journal} {J. Chem. Phys.}\ }\textbf
  {\bibinfo {volume} {138}},\ \bibinfo {pages} {03B603} (\bibinfo {year}
  {2013})}\BibitemShut {NoStop}%
\bibitem [{\citenamefont {White}\ and\ \citenamefont
  {Voth}(2014)}]{white2014efficient}%
  \BibitemOpen
  \bibfield  {author} {\bibinfo {author} {\bibfnamefont {A.~D.}\ \bibnamefont
  {White}}\ and\ \bibinfo {author} {\bibfnamefont {G.~A.}\ \bibnamefont
  {Voth}},\ }\bibfield  {title} {\enquote {\bibinfo {title} {Efficient and
  minimal method to bias molecular simulations with experimental data},}\
  }\href@noop {} {\bibfield  {journal} {\bibinfo  {journal} {J. Chem. Theory
  Comput.}\ }\textbf {\bibinfo {volume} {10}},\ \bibinfo {pages} {3023--3030}
  (\bibinfo {year} {2014})}\BibitemShut {NoStop}%
\bibitem [{\citenamefont {Hummer}\ and\ \citenamefont
  {K{\"o}finger}(2015)}]{hummer2015bayesian}%
  \BibitemOpen
  \bibfield  {author} {\bibinfo {author} {\bibfnamefont {G.}~\bibnamefont
  {Hummer}}\ and\ \bibinfo {author} {\bibfnamefont {J.}~\bibnamefont
  {K{\"o}finger}},\ }\bibfield  {title} {\enquote {\bibinfo {title} {Bayesian
  ensemble refinement by replica simulations and reweighting},}\ }\href@noop {}
  {\bibfield  {journal} {\bibinfo  {journal} {J. Chem. Phys.}\ }\textbf
  {\bibinfo {volume} {143}},\ \bibinfo {pages} {12B634\_1} (\bibinfo {year}
  {2015})}\BibitemShut {NoStop}%
\bibitem [{\citenamefont {Bonomi}\ \emph {et~al.}(2016)\citenamefont {Bonomi},
  \citenamefont {Camilloni}, \citenamefont {Cavalli},\ and\ \citenamefont
  {Vendruscolo}}]{bonomi2016metainference}%
  \BibitemOpen
  \bibfield  {author} {\bibinfo {author} {\bibfnamefont {M.}~\bibnamefont
  {Bonomi}}, \bibinfo {author} {\bibfnamefont {C.}~\bibnamefont {Camilloni}},
  \bibinfo {author} {\bibfnamefont {A.}~\bibnamefont {Cavalli}}, \ and\
  \bibinfo {author} {\bibfnamefont {M.}~\bibnamefont {Vendruscolo}},\
  }\bibfield  {title} {\enquote {\bibinfo {title} {Metainference: A {B}ayesian
  inference method for heterogeneous systems},}\ }\href@noop {} {\bibfield
  {journal} {\bibinfo  {journal} {Sci. Adv.}\ }\textbf {\bibinfo {volume}
  {2}},\ \bibinfo {pages} {e1501177} (\bibinfo {year} {2016})}\BibitemShut
  {NoStop}%
\bibitem [{\citenamefont {Costa}\ and\ \citenamefont
  {Fushman}(2022)}]{costa2022reweighting}%
  \BibitemOpen
  \bibfield  {author} {\bibinfo {author} {\bibfnamefont {R.~G.~L.}\
  \bibnamefont {Costa}}\ and\ \bibinfo {author} {\bibfnamefont
  {D.}~\bibnamefont {Fushman}},\ }\bibfield  {title} {\enquote {\bibinfo
  {title} {Reweighting methods for elucidation of conformation ensembles of
  proteins},}\ }\href@noop {} {\bibfield  {journal} {\bibinfo  {journal} {Curr.
  Opin. Struct. Biol.}\ }\textbf {\bibinfo {volume} {77}},\ \bibinfo {pages}
  {102470} (\bibinfo {year} {2022})}\BibitemShut {NoStop}%
\bibitem [{\citenamefont {Bernad{\'o}}\ \emph {et~al.}(2007)\citenamefont
  {Bernad{\'o}}, \citenamefont {Mylonas}, \citenamefont {Petoukhov},
  \citenamefont {Blackledge},\ and\ \citenamefont
  {Svergun}}]{bernado2007structural}%
  \BibitemOpen
  \bibfield  {author} {\bibinfo {author} {\bibfnamefont {P.}~\bibnamefont
  {Bernad{\'o}}}, \bibinfo {author} {\bibfnamefont {E.}~\bibnamefont
  {Mylonas}}, \bibinfo {author} {\bibfnamefont {M.~V.}\ \bibnamefont
  {Petoukhov}}, \bibinfo {author} {\bibfnamefont {M.}~\bibnamefont
  {Blackledge}}, \ and\ \bibinfo {author} {\bibfnamefont {D.~I.}\ \bibnamefont
  {Svergun}},\ }\bibfield  {title} {\enquote {\bibinfo {title} {Structural
  characterization of flexible proteins using small-angle {X}-ray
  scattering},}\ }\href@noop {} {\bibfield  {journal} {\bibinfo  {journal} {J.
  Am. Chem. Soc.}\ }\textbf {\bibinfo {volume} {129}},\ \bibinfo {pages}
  {5656--5664} (\bibinfo {year} {2007})}\BibitemShut {NoStop}%
\bibitem [{\citenamefont {Tria}\ \emph {et~al.}(2015)\citenamefont {Tria},
  \citenamefont {Mertens}, \citenamefont {Kachala},\ and\ \citenamefont
  {Svergun}}]{tria2015advanced}%
  \BibitemOpen
  \bibfield  {author} {\bibinfo {author} {\bibfnamefont {G.}~\bibnamefont
  {Tria}}, \bibinfo {author} {\bibfnamefont {H.~D.}\ \bibnamefont {Mertens}},
  \bibinfo {author} {\bibfnamefont {M.}~\bibnamefont {Kachala}}, \ and\
  \bibinfo {author} {\bibfnamefont {D.~I.}\ \bibnamefont {Svergun}},\
  }\bibfield  {title} {\enquote {\bibinfo {title} {Advanced ensemble modelling
  of flexible macromolecules using {X}-ray solution scattering},}\ }\href@noop
  {} {\bibfield  {journal} {\bibinfo  {journal} {IUCrJ}\ }\textbf {\bibinfo
  {volume} {2}},\ \bibinfo {pages} {207--217} (\bibinfo {year}
  {2015})}\BibitemShut {NoStop}%
\bibitem [{\citenamefont {Pitera}\ and\ \citenamefont
  {Chodera}(2012)}]{pitera2012use}%
  \BibitemOpen
  \bibfield  {author} {\bibinfo {author} {\bibfnamefont {J.~W.}\ \bibnamefont
  {Pitera}}\ and\ \bibinfo {author} {\bibfnamefont {J.~D.}\ \bibnamefont
  {Chodera}},\ }\bibfield  {title} {\enquote {\bibinfo {title} {On the use of
  experimental observations to bias simulated ensembles},}\ }\href@noop {}
  {\bibfield  {journal} {\bibinfo  {journal} {J. Chem. Theory Comput.}\
  }\textbf {\bibinfo {volume} {8}},\ \bibinfo {pages} {3445--3451} (\bibinfo
  {year} {2012})}\BibitemShut {NoStop}%
\bibitem [{\citenamefont {Brookes}\ and\ \citenamefont
  {Head-Gordon}(2016)}]{brookes2016experimental}%
  \BibitemOpen
  \bibfield  {author} {\bibinfo {author} {\bibfnamefont {D.~H.}\ \bibnamefont
  {Brookes}}\ and\ \bibinfo {author} {\bibfnamefont {T.}~\bibnamefont
  {Head-Gordon}},\ }\bibfield  {title} {\enquote {\bibinfo {title}
  {Experimental inferential structure determination of ensembles for
  intrinsically disordered proteins},}\ }\href@noop {} {\bibfield  {journal}
  {\bibinfo  {journal} {J. Am. Chem. Soc.}\ }\textbf {\bibinfo {volume}
  {138}},\ \bibinfo {pages} {4530--4538} (\bibinfo {year} {2016})}\BibitemShut
  {NoStop}%
\bibitem [{\citenamefont {K{\"o}finger}\ \emph {et~al.}(2019)\citenamefont
  {K{\"o}finger}, \citenamefont {Stelzl}, \citenamefont {Reuter}, \citenamefont
  {Allande}, \citenamefont {Reichel},\ and\ \citenamefont
  {Hummer}}]{kofinger2019efficient}%
  \BibitemOpen
  \bibfield  {author} {\bibinfo {author} {\bibfnamefont {J.}~\bibnamefont
  {K{\"o}finger}}, \bibinfo {author} {\bibfnamefont {L.~S.}\ \bibnamefont
  {Stelzl}}, \bibinfo {author} {\bibfnamefont {K.}~\bibnamefont {Reuter}},
  \bibinfo {author} {\bibfnamefont {C.}~\bibnamefont {Allande}}, \bibinfo
  {author} {\bibfnamefont {K.}~\bibnamefont {Reichel}}, \ and\ \bibinfo
  {author} {\bibfnamefont {G.}~\bibnamefont {Hummer}},\ }\bibfield  {title}
  {\enquote {\bibinfo {title} {Efficient ensemble refinement by reweighting},}\
  }\href@noop {} {\bibfield  {journal} {\bibinfo  {journal} {J. Chem. Theory
  Comput.}\ }\textbf {\bibinfo {volume} {15}},\ \bibinfo {pages} {3390--3401}
  (\bibinfo {year} {2019})}\BibitemShut {NoStop}%
\bibitem [{\citenamefont {Bottaro}, \citenamefont {Bengtsen},\ and\
  \citenamefont {Lindorff-Larsen}(2020)}]{bottaro2020integrating}%
  \BibitemOpen
  \bibfield  {author} {\bibinfo {author} {\bibfnamefont {S.}~\bibnamefont
  {Bottaro}}, \bibinfo {author} {\bibfnamefont {T.}~\bibnamefont {Bengtsen}}, \
  and\ \bibinfo {author} {\bibfnamefont {K.}~\bibnamefont {Lindorff-Larsen}},\
  }\bibfield  {title} {\enquote {\bibinfo {title} {Integrating molecular
  simulation and experimental data: a {B}ayesian/maximum entropy reweighting
  approach},}\ }\href@noop {} {\bibfield  {journal} {\bibinfo  {journal}
  {Structural bioinformatics: methods and protocols}\ ,\ \bibinfo {pages}
  {219--240}} (\bibinfo {year} {2020})}\BibitemShut {NoStop}%
\bibitem [{\citenamefont {Medeiros~Selegato}\ \emph {et~al.}(2021)\citenamefont
  {Medeiros~Selegato}, \citenamefont {Bracco}, \citenamefont {Giannelli},
  \citenamefont {Parigi}, \citenamefont {Luchinat}, \citenamefont {Sgheri},\
  and\ \citenamefont {Ravera}}]{medeiros2021comparison}%
  \BibitemOpen
  \bibfield  {author} {\bibinfo {author} {\bibfnamefont {D.}~\bibnamefont
  {Medeiros~Selegato}}, \bibinfo {author} {\bibfnamefont {C.}~\bibnamefont
  {Bracco}}, \bibinfo {author} {\bibfnamefont {C.}~\bibnamefont {Giannelli}},
  \bibinfo {author} {\bibfnamefont {G.}~\bibnamefont {Parigi}}, \bibinfo
  {author} {\bibfnamefont {C.}~\bibnamefont {Luchinat}}, \bibinfo {author}
  {\bibfnamefont {L.}~\bibnamefont {Sgheri}}, \ and\ \bibinfo {author}
  {\bibfnamefont {E.}~\bibnamefont {Ravera}},\ }\bibfield  {title} {\enquote
  {\bibinfo {title} {Comparison of different reweighting approaches for the
  calculation of conformational variability of macromolecules from molecular
  simulations},}\ }\href@noop {} {\bibfield  {journal} {\bibinfo  {journal}
  {ChemPhysChem}\ }\textbf {\bibinfo {volume} {22}},\ \bibinfo {pages}
  {127--138} (\bibinfo {year} {2021})}\BibitemShut {NoStop}%
\bibitem [{\citenamefont {Cesari}, \citenamefont {Rei{\ss}er},\ and\
  \citenamefont {Bussi}(2018)}]{cesari2018using}%
  \BibitemOpen
  \bibfield  {author} {\bibinfo {author} {\bibfnamefont {A.}~\bibnamefont
  {Cesari}}, \bibinfo {author} {\bibfnamefont {S.}~\bibnamefont {Rei{\ss}er}},
  \ and\ \bibinfo {author} {\bibfnamefont {G.}~\bibnamefont {Bussi}},\
  }\bibfield  {title} {\enquote {\bibinfo {title} {Using the maximum entropy
  principle to combine simulations and solution experiments},}\ }\href@noop {}
  {\bibfield  {journal} {\bibinfo  {journal} {Computation}\ }\textbf {\bibinfo
  {volume} {6}},\ \bibinfo {pages} {15} (\bibinfo {year} {2018})}\BibitemShut
  {NoStop}%
\bibitem [{\citenamefont {Svergun}, \citenamefont {Barberato},\ and\
  \citenamefont {Koch}(1995)}]{svergun1995crysol}%
  \BibitemOpen
  \bibfield  {author} {\bibinfo {author} {\bibfnamefont {D.}~\bibnamefont
  {Svergun}}, \bibinfo {author} {\bibfnamefont {C.}~\bibnamefont {Barberato}},
  \ and\ \bibinfo {author} {\bibfnamefont {M.~H.}\ \bibnamefont {Koch}},\
  }\bibfield  {title} {\enquote {\bibinfo {title} {{CRYSOL}--a program to
  evaluate {X}-ray solution scattering of biological macromolecules from atomic
  coordinates},}\ }\href@noop {} {\bibfield  {journal} {\bibinfo  {journal}
  {Journal of applied crystallography}\ }\textbf {\bibinfo {volume} {28}},\
  \bibinfo {pages} {768--773} (\bibinfo {year} {1995})}\BibitemShut {NoStop}%
\bibitem [{\citenamefont {K{\"o}finger}\ and\ \citenamefont
  {Hummer}(2013)}]{kofinger2013atomic}%
  \BibitemOpen
  \bibfield  {author} {\bibinfo {author} {\bibfnamefont {J.}~\bibnamefont
  {K{\"o}finger}}\ and\ \bibinfo {author} {\bibfnamefont {G.}~\bibnamefont
  {Hummer}},\ }\bibfield  {title} {\enquote {\bibinfo {title}
  {Atomic-resolution structural information from scattering experiments on
  macromolecules in solution},}\ }\href@noop {} {\bibfield  {journal} {\bibinfo
   {journal} {Phy. Rev. E}\ }\textbf {\bibinfo {volume} {87}},\ \bibinfo
  {pages} {052712} (\bibinfo {year} {2013})}\BibitemShut {NoStop}%
\bibitem [{\citenamefont {Knight}\ and\ \citenamefont
  {Hub}(2015)}]{knight2015waxsis}%
  \BibitemOpen
  \bibfield  {author} {\bibinfo {author} {\bibfnamefont {C.~J.}\ \bibnamefont
  {Knight}}\ and\ \bibinfo {author} {\bibfnamefont {J.~S.}\ \bibnamefont
  {Hub}},\ }\bibfield  {title} {\enquote {\bibinfo {title} {Waxsis: a web
  server for the calculation of saxs/waxs curves based on explicit-solvent
  molecular dynamics},}\ }\href@noop {} {\bibfield  {journal} {\bibinfo
  {journal} {Nucleic Acids Res.}\ }\textbf {\bibinfo {volume} {43}},\ \bibinfo
  {pages} {W225--W230} (\bibinfo {year} {2015})}\BibitemShut {NoStop}%
\bibitem [{\citenamefont {Bonomi}\ \emph {et~al.}(2019)\citenamefont {Bonomi},
  \citenamefont {Hanot}, \citenamefont {Greenberg}, \citenamefont {Sali},
  \citenamefont {Nilges}, \citenamefont {Vendruscolo},\ and\ \citenamefont
  {Pellarin}}]{bonomi2019bayesian}%
  \BibitemOpen
  \bibfield  {author} {\bibinfo {author} {\bibfnamefont {M.}~\bibnamefont
  {Bonomi}}, \bibinfo {author} {\bibfnamefont {S.}~\bibnamefont {Hanot}},
  \bibinfo {author} {\bibfnamefont {C.~H.}\ \bibnamefont {Greenberg}}, \bibinfo
  {author} {\bibfnamefont {A.}~\bibnamefont {Sali}}, \bibinfo {author}
  {\bibfnamefont {M.}~\bibnamefont {Nilges}}, \bibinfo {author} {\bibfnamefont
  {M.}~\bibnamefont {Vendruscolo}}, \ and\ \bibinfo {author} {\bibfnamefont
  {R.}~\bibnamefont {Pellarin}},\ }\bibfield  {title} {\enquote {\bibinfo
  {title} {Bayesian weighing of electron cryo-microscopy data for integrative
  structural modeling},}\ }\href@noop {} {\bibfield  {journal} {\bibinfo
  {journal} {Structure}\ }\textbf {\bibinfo {volume} {27}},\ \bibinfo {pages}
  {175--188} (\bibinfo {year} {2019})}\BibitemShut {NoStop}%
\bibitem [{\citenamefont {Karplus}(1963)}]{karplus1963vicinal}%
  \BibitemOpen
  \bibfield  {author} {\bibinfo {author} {\bibfnamefont {M.}~\bibnamefont
  {Karplus}},\ }\bibfield  {title} {\enquote {\bibinfo {title} {Vicinal proton
  coupling in nuclear magnetic resonance},}\ }\href@noop {} {\bibfield
  {journal} {\bibinfo  {journal} {J. Am. Chem. Soc.}\ }\textbf {\bibinfo
  {volume} {85}},\ \bibinfo {pages} {2870--2871} (\bibinfo {year}
  {1963})}\BibitemShut {NoStop}%
\bibitem [{\citenamefont {Morales}\ and\ \citenamefont
  {Nocedal}(2011)}]{morales2011remark}%
  \BibitemOpen
  \bibfield  {author} {\bibinfo {author} {\bibfnamefont {J.~L.}\ \bibnamefont
  {Morales}}\ and\ \bibinfo {author} {\bibfnamefont {J.}~\bibnamefont
  {Nocedal}},\ }\bibfield  {title} {\enquote {\bibinfo {title} {Remark on
  “algorithm 778: {L-BFGS-B}: Fortran subroutines for large-scale bound
  constrained optimization”},}\ }\href@noop {} {\bibfield  {journal}
  {\bibinfo  {journal} {ACM Transactions on Mathematical Software (TOMS)}\
  }\textbf {\bibinfo {volume} {38}},\ \bibinfo {pages} {1--4} (\bibinfo {year}
  {2011})}\BibitemShut {NoStop}%
\bibitem [{\citenamefont {Virtanen}\ \emph {et~al.}(2020)\citenamefont
  {Virtanen}, \citenamefont {Gommers}, \citenamefont {Oliphant}, \citenamefont
  {Haberland}, \citenamefont {Reddy}, \citenamefont {Cournapeau}, \citenamefont
  {Burovski}, \citenamefont {Peterson}, \citenamefont {Weckesser},
  \citenamefont {Bright} \emph {et~al.}}]{virtanen2020scipy}%
  \BibitemOpen
  \bibfield  {author} {\bibinfo {author} {\bibfnamefont {P.}~\bibnamefont
  {Virtanen}}, \bibinfo {author} {\bibfnamefont {R.}~\bibnamefont {Gommers}},
  \bibinfo {author} {\bibfnamefont {T.~E.}\ \bibnamefont {Oliphant}}, \bibinfo
  {author} {\bibfnamefont {M.}~\bibnamefont {Haberland}}, \bibinfo {author}
  {\bibfnamefont {T.}~\bibnamefont {Reddy}}, \bibinfo {author} {\bibfnamefont
  {D.}~\bibnamefont {Cournapeau}}, \bibinfo {author} {\bibfnamefont
  {E.}~\bibnamefont {Burovski}}, \bibinfo {author} {\bibfnamefont
  {P.}~\bibnamefont {Peterson}}, \bibinfo {author} {\bibfnamefont
  {W.}~\bibnamefont {Weckesser}}, \bibinfo {author} {\bibfnamefont
  {J.}~\bibnamefont {Bright}},  \emph {et~al.},\ }\bibfield  {title} {\enquote
  {\bibinfo {title} {Scipy 1.0: fundamental algorithms for scientific computing
  in python},}\ }\href@noop {} {\bibfield  {journal} {\bibinfo  {journal} {Nat.
  Methods}\ }\textbf {\bibinfo {volume} {17}},\ \bibinfo {pages} {261--272}
  (\bibinfo {year} {2020})}\BibitemShut {NoStop}%
\bibitem [{\citenamefont {Fröhlking}\ \emph {et~al.}(2020)\citenamefont
  {Fröhlking}, \citenamefont {Bernetti}, \citenamefont {Calonaci},\ and\
  \citenamefont {Bussi}}]{froehlking2020toward}%
  \BibitemOpen
  \bibfield  {author} {\bibinfo {author} {\bibfnamefont {T.}~\bibnamefont
  {Fröhlking}}, \bibinfo {author} {\bibfnamefont {M.}~\bibnamefont
  {Bernetti}}, \bibinfo {author} {\bibfnamefont {N.}~\bibnamefont {Calonaci}},
  \ and\ \bibinfo {author} {\bibfnamefont {G.}~\bibnamefont {Bussi}},\
  }\bibfield  {title} {\enquote {\bibinfo {title} {Toward empirical force
  fields that match experimental observables},}\ }\href@noop {} {\bibfield
  {journal} {\bibinfo  {journal} {J. Chem. Phys.}\ }\textbf {\bibinfo {volume}
  {152}},\ \bibinfo {pages} {230902} (\bibinfo {year} {2020})}\BibitemShut
  {NoStop}%
\bibitem [{\citenamefont {Cornell}\ \emph {et~al.}(1996)\citenamefont
  {Cornell}, \citenamefont {Cieplak}, \citenamefont {Bayly}, \citenamefont
  {Gould}, \citenamefont {Merz}, \citenamefont {Ferguson}, \citenamefont
  {Spellmeyer}, \citenamefont {Fox}, \citenamefont {Caldwell},\ and\
  \citenamefont {Kollman}}]{Cornell1996}%
  \BibitemOpen
  \bibfield  {author} {\bibinfo {author} {\bibfnamefont {W.~D.}\ \bibnamefont
  {Cornell}}, \bibinfo {author} {\bibfnamefont {P.}~\bibnamefont {Cieplak}},
  \bibinfo {author} {\bibfnamefont {C.~I.}\ \bibnamefont {Bayly}}, \bibinfo
  {author} {\bibfnamefont {I.~R.}\ \bibnamefont {Gould}}, \bibinfo {author}
  {\bibfnamefont {K.~M.}\ \bibnamefont {Merz}}, \bibinfo {author}
  {\bibfnamefont {D.~M.}\ \bibnamefont {Ferguson}}, \bibinfo {author}
  {\bibfnamefont {D.~C.}\ \bibnamefont {Spellmeyer}}, \bibinfo {author}
  {\bibfnamefont {T.}~\bibnamefont {Fox}}, \bibinfo {author} {\bibfnamefont
  {J.~W.}\ \bibnamefont {Caldwell}}, \ and\ \bibinfo {author} {\bibfnamefont
  {P.~A.}\ \bibnamefont {Kollman}},\ }\bibfield  {title} {\enquote {\bibinfo
  {title} {A second generation force field for the simulation of proteins,
  nucleic acids, and organic molecules},}\ }\href@noop {} {\bibfield  {journal}
  {\bibinfo  {journal} {J. Am. Chem. Soc.}\ }\textbf {\bibinfo {volume}
  {118}},\ \bibinfo {pages} {2309--2309} (\bibinfo {year} {1996})}\BibitemShut
  {NoStop}%
\bibitem [{\citenamefont {Wang}, \citenamefont {Cieplak},\ and\ \citenamefont
  {Kollman}(2000)}]{Wang2000}%
  \BibitemOpen
  \bibfield  {author} {\bibinfo {author} {\bibfnamefont {J.}~\bibnamefont
  {Wang}}, \bibinfo {author} {\bibfnamefont {P.}~\bibnamefont {Cieplak}}, \
  and\ \bibinfo {author} {\bibfnamefont {P.~A.}\ \bibnamefont {Kollman}},\
  }\bibfield  {title} {\enquote {\bibinfo {title} {How well does a restrained
  electrostatic potential ({RESP}) model perform in calculating conformational
  energies of organic and biological molecules?}}\ }\href@noop {} {\bibfield
  {journal} {\bibinfo  {journal} {J. Comput. Chem.}\ }\textbf {\bibinfo
  {volume} {21}},\ \bibinfo {pages} {1049--1074} (\bibinfo {year}
  {2000})}\BibitemShut {NoStop}%
\bibitem [{\citenamefont {Pérez}\ \emph {et~al.}(2007)\citenamefont {Pérez},
  \citenamefont {Marchán}, \citenamefont {Svozil}, \citenamefont {Sponer},
  \citenamefont {Cheatham}, \citenamefont {Laughton},\ and\ \citenamefont
  {Orozco}}]{Perez2007}%
  \BibitemOpen
  \bibfield  {author} {\bibinfo {author} {\bibfnamefont {A.}~\bibnamefont
  {Pérez}}, \bibinfo {author} {\bibfnamefont {I.}~\bibnamefont {Marchán}},
  \bibinfo {author} {\bibfnamefont {D.}~\bibnamefont {Svozil}}, \bibinfo
  {author} {\bibfnamefont {J.}~\bibnamefont {Sponer}}, \bibinfo {author}
  {\bibfnamefont {T.~E.}\ \bibnamefont {Cheatham}}, \bibinfo {author}
  {\bibfnamefont {C.~A.}\ \bibnamefont {Laughton}}, \ and\ \bibinfo {author}
  {\bibfnamefont {M.}~\bibnamefont {Orozco}},\ }\bibfield  {title} {\enquote
  {\bibinfo {title} {Refinement of the {AMBER} force field for nucleic acids:
  Improving the description of $\alpha$/$\gamma$ conformers},}\ }\href@noop {}
  {\bibfield  {journal} {\bibinfo  {journal} {Biophys. J.}\ }\textbf {\bibinfo
  {volume} {92}},\ \bibinfo {pages} {3817--3829} (\bibinfo {year}
  {2007})}\BibitemShut {NoStop}%
\bibitem [{\citenamefont {Zgarbov\'a}\ \emph {et~al.}(2011)\citenamefont
  {Zgarbov\'a}, \citenamefont {Otyepka}, \citenamefont {\v{S}poner},
  \citenamefont {Ml\'{a}dek}, \citenamefont {Ban\'{a}\v{s}}, \citenamefont
  {Cheatham},\ and\ \citenamefont {Jure\v{c}ka}}]{Zgarbova2011}%
  \BibitemOpen
  \bibfield  {author} {\bibinfo {author} {\bibfnamefont {M.}~\bibnamefont
  {Zgarbov\'a}}, \bibinfo {author} {\bibfnamefont {M.}~\bibnamefont {Otyepka}},
  \bibinfo {author} {\bibfnamefont {J.}~\bibnamefont {\v{S}poner}}, \bibinfo
  {author} {\bibfnamefont {A.}~\bibnamefont {Ml\'{a}dek}}, \bibinfo {author}
  {\bibfnamefont {P.}~\bibnamefont {Ban\'{a}\v{s}}}, \bibinfo {author}
  {\bibfnamefont {T.~E.}\ \bibnamefont {Cheatham}}, \ and\ \bibinfo {author}
  {\bibfnamefont {P.}~\bibnamefont {Jure\v{c}ka}},\ }\bibfield  {title}
  {\enquote {\bibinfo {title} {Refinement of the {C}ornell et al. nucleic acids
  force field based on reference quantum chemical calculations of glycosidic
  torsion profiles},}\ }\href@noop {} {\bibfield  {journal} {\bibinfo
  {journal} {J. Chem. Theory Comput.}\ }\textbf {\bibinfo {volume} {7}},\
  \bibinfo {pages} {2886--2902} (\bibinfo {year} {2011})}\BibitemShut {NoStop}%
\bibitem [{\citenamefont {Steinbrecher}, \citenamefont {Latzer},\ and\
  \citenamefont {Case}(2012)}]{Steinbrecher2012}%
  \BibitemOpen
  \bibfield  {author} {\bibinfo {author} {\bibfnamefont {T.}~\bibnamefont
  {Steinbrecher}}, \bibinfo {author} {\bibfnamefont {J.}~\bibnamefont
  {Latzer}}, \ and\ \bibinfo {author} {\bibfnamefont {D.~A.}\ \bibnamefont
  {Case}},\ }\bibfield  {title} {\enquote {\bibinfo {title} {Revised {AMBER}
  parameters for bioorganic phosphates},}\ }\href@noop {} {\bibfield  {journal}
  {\bibinfo  {journal} {J. Chem. Theory Comput.}\ }\textbf {\bibinfo {volume}
  {8}},\ \bibinfo {pages} {4405--4412} (\bibinfo {year} {2012})}\BibitemShut
  {NoStop}%
\bibitem [{\citenamefont {Izadi}, \citenamefont {Anandakrishnan},\ and\
  \citenamefont {Onufriev}(2014)}]{IzadiOPC2014}%
  \BibitemOpen
  \bibfield  {author} {\bibinfo {author} {\bibfnamefont {S.}~\bibnamefont
  {Izadi}}, \bibinfo {author} {\bibfnamefont {R.}~\bibnamefont
  {Anandakrishnan}}, \ and\ \bibinfo {author} {\bibfnamefont {A.~V.}\
  \bibnamefont {Onufriev}},\ }\bibfield  {title} {\enquote {\bibinfo {title}
  {Building water models: A different approach},}\ }\href@noop {} {\bibfield
  {journal} {\bibinfo  {journal} {J. Phys. Chem. Lett.}\ }\textbf {\bibinfo
  {volume} {5}},\ \bibinfo {pages} {3863--3871} (\bibinfo {year}
  {2014})}\BibitemShut {NoStop}%
\bibitem [{\citenamefont {Bergonzo}\ and\ \citenamefont
  {Cheatham~III}(2015)}]{bergonzo2015improved}%
  \BibitemOpen
  \bibfield  {author} {\bibinfo {author} {\bibfnamefont {C.}~\bibnamefont
  {Bergonzo}}\ and\ \bibinfo {author} {\bibfnamefont {T.~E.}\ \bibnamefont
  {Cheatham~III}},\ }\bibfield  {title} {\enquote {\bibinfo {title} {Improved
  force field parameters lead to a better description of {RNA} structure},}\
  }\href@noop {} {\bibfield  {journal} {\bibinfo  {journal} {J. Chem. Theory
  Comput.}\ }\textbf {\bibinfo {volume} {11}},\ \bibinfo {pages} {3969--3972}
  (\bibinfo {year} {2015})}\BibitemShut {NoStop}%
\bibitem [{\citenamefont {Bergonzo}, \citenamefont {Grishaev},\ and\
  \citenamefont {Bottaro}(2022)}]{bergonzo2022conformational}%
  \BibitemOpen
  \bibfield  {author} {\bibinfo {author} {\bibfnamefont {C.}~\bibnamefont
  {Bergonzo}}, \bibinfo {author} {\bibfnamefont {A.~V.}\ \bibnamefont
  {Grishaev}}, \ and\ \bibinfo {author} {\bibfnamefont {S.}~\bibnamefont
  {Bottaro}},\ }\bibfield  {title} {\enquote {\bibinfo {title} {Conformational
  heterogeneity of {UCAAUC} {RNA} oligonucleotide from molecular dynamics
  simulations, {SAXS}, and {NMR} experiments},}\ }\href@noop {} {\bibfield
  {journal} {\bibinfo  {journal} {RNA}\ }\textbf {\bibinfo {volume} {28}},\
  \bibinfo {pages} {937--946} (\bibinfo {year} {2022})}\BibitemShut {NoStop}%
\bibitem [{\citenamefont {Joung}\ and\ \citenamefont
  {Cheatham}(2008)}]{CheathamIons2008}%
  \BibitemOpen
  \bibfield  {author} {\bibinfo {author} {\bibfnamefont {I.~S.}\ \bibnamefont
  {Joung}}\ and\ \bibinfo {author} {\bibfnamefont {T.~E.}\ \bibnamefont
  {Cheatham}},\ }\bibfield  {title} {\enquote {\bibinfo {title} {Determination
  of alkali and halide monovalent ion parameters for use in explicitly solvated
  biomolecular simulations},}\ }\href@noop {} {\bibfield  {journal} {\bibinfo
  {journal} {J. Phys. Chem. B}\ }\textbf {\bibinfo {volume} {112}},\ \bibinfo
  {pages} {9020--9041} (\bibinfo {year} {2008})}\BibitemShut {NoStop}%
\bibitem [{\citenamefont {Abraham}\ \emph {et~al.}(2015)\citenamefont
  {Abraham}, \citenamefont {Murtola}, \citenamefont {Schulz}, \citenamefont
  {P{\'a}ll}, \citenamefont {Smith}, \citenamefont {Hess},\ and\ \citenamefont
  {Lindahl}}]{abraham2015gromacs}%
  \BibitemOpen
  \bibfield  {author} {\bibinfo {author} {\bibfnamefont {M.~J.}\ \bibnamefont
  {Abraham}}, \bibinfo {author} {\bibfnamefont {T.}~\bibnamefont {Murtola}},
  \bibinfo {author} {\bibfnamefont {R.}~\bibnamefont {Schulz}}, \bibinfo
  {author} {\bibfnamefont {S.}~\bibnamefont {P{\'a}ll}}, \bibinfo {author}
  {\bibfnamefont {J.~C.}\ \bibnamefont {Smith}}, \bibinfo {author}
  {\bibfnamefont {B.}~\bibnamefont {Hess}}, \ and\ \bibinfo {author}
  {\bibfnamefont {E.}~\bibnamefont {Lindahl}},\ }\bibfield  {title} {\enquote
  {\bibinfo {title} {{GROMACS}: High performance molecular simulations through
  multi-level parallelism from laptops to supercomputers},}\ }\href@noop {}
  {\bibfield  {journal} {\bibinfo  {journal} {SoftwareX}\ }\textbf {\bibinfo
  {volume} {1}},\ \bibinfo {pages} {19--25} (\bibinfo {year}
  {2015})}\BibitemShut {NoStop}%
\bibitem [{\citenamefont {Hansmann}(1997)}]{hansmann1997}%
  \BibitemOpen
  \bibfield  {author} {\bibinfo {author} {\bibfnamefont {U.~H.}\ \bibnamefont
  {Hansmann}},\ }\bibfield  {title} {\enquote {\bibinfo {title} {Parallel
  tempering algorithm for conformational studies of biological molecules},}\
  }\href {\doibase https://doi.org/10.1016/S0009-2614(97)01198-6} {\bibfield
  {journal} {\bibinfo  {journal} {Chem. Phys. Lett.}\ }\textbf {\bibinfo
  {volume} {281}},\ \bibinfo {pages} {140--150} (\bibinfo {year}
  {1997})}\BibitemShut {NoStop}%
\bibitem [{\citenamefont {Sugita}\ and\ \citenamefont
  {Okamoto}(1999)}]{sugita1999}%
  \BibitemOpen
  \bibfield  {author} {\bibinfo {author} {\bibfnamefont {Y.}~\bibnamefont
  {Sugita}}\ and\ \bibinfo {author} {\bibfnamefont {Y.}~\bibnamefont
  {Okamoto}},\ }\bibfield  {title} {\enquote {\bibinfo {title}
  {Replica-exchange molecular dynamics method for protein folding},}\ }\href
  {\doibase https://doi.org/10.1016/S0009-2614(99)01123-9} {\bibfield
  {journal} {\bibinfo  {journal} {Chem. Phys. Lett.}\ }\textbf {\bibinfo
  {volume} {314}},\ \bibinfo {pages} {141--151} (\bibinfo {year}
  {1999})}\BibitemShut {NoStop}%
\bibitem [{\citenamefont {Condon}\ \emph {et~al.}(2015)\citenamefont {Condon},
  \citenamefont {Kennedy}, \citenamefont {Mort}, \citenamefont {Kierzek},
  \citenamefont {Yildirim},\ and\ \citenamefont {Turner}}]{condon2015stacking}%
  \BibitemOpen
  \bibfield  {author} {\bibinfo {author} {\bibfnamefont {D.~E.}\ \bibnamefont
  {Condon}}, \bibinfo {author} {\bibfnamefont {S.~D.}\ \bibnamefont {Kennedy}},
  \bibinfo {author} {\bibfnamefont {B.~C.}\ \bibnamefont {Mort}}, \bibinfo
  {author} {\bibfnamefont {R.}~\bibnamefont {Kierzek}}, \bibinfo {author}
  {\bibfnamefont {I.}~\bibnamefont {Yildirim}}, \ and\ \bibinfo {author}
  {\bibfnamefont {D.~H.}\ \bibnamefont {Turner}},\ }\bibfield  {title}
  {\enquote {\bibinfo {title} {Stacking in {RNA}: {NMR} of four tetramers
  benchmark molecular dynamics},}\ }\href@noop {} {\bibfield  {journal}
  {\bibinfo  {journal} {J. Chem. Theory Comput.}\ }\textbf {\bibinfo {volume}
  {11}},\ \bibinfo {pages} {2729--2742} (\bibinfo {year} {2015})}\BibitemShut
  {NoStop}%
\bibitem [{\citenamefont {Tubbs}\ \emph {et~al.}(2013)\citenamefont {Tubbs},
  \citenamefont {Condon}, \citenamefont {Kennedy}, \citenamefont {Hauser},
  \citenamefont {Bevilacqua},\ and\ \citenamefont {Turner}}]{tubbs2013nuclear}%
  \BibitemOpen
  \bibfield  {author} {\bibinfo {author} {\bibfnamefont {J.~D.}\ \bibnamefont
  {Tubbs}}, \bibinfo {author} {\bibfnamefont {D.~E.}\ \bibnamefont {Condon}},
  \bibinfo {author} {\bibfnamefont {S.~D.}\ \bibnamefont {Kennedy}}, \bibinfo
  {author} {\bibfnamefont {M.}~\bibnamefont {Hauser}}, \bibinfo {author}
  {\bibfnamefont {P.~C.}\ \bibnamefont {Bevilacqua}}, \ and\ \bibinfo {author}
  {\bibfnamefont {D.~H.}\ \bibnamefont {Turner}},\ }\bibfield  {title}
  {\enquote {\bibinfo {title} {The nuclear magnetic resonance of {CCCC} {RNA}
  reveals a right-handed helix, and revised parameters for {AMBER} force field
  torsions improve structural predictions from molecular dynamics},}\
  }\href@noop {} {\bibfield  {journal} {\bibinfo  {journal} {Biochemistry}\
  }\textbf {\bibinfo {volume} {52}},\ \bibinfo {pages} {996--1010} (\bibinfo
  {year} {2013})}\BibitemShut {NoStop}%
\bibitem [{\citenamefont {Yildirim}\ \emph {et~al.}(2011)\citenamefont
  {Yildirim}, \citenamefont {Stern}, \citenamefont {Tubbs}, \citenamefont
  {Kennedy},\ and\ \citenamefont {Turner}}]{yildirim2011benchmarking}%
  \BibitemOpen
  \bibfield  {author} {\bibinfo {author} {\bibfnamefont {I.}~\bibnamefont
  {Yildirim}}, \bibinfo {author} {\bibfnamefont {H.~A.}\ \bibnamefont {Stern}},
  \bibinfo {author} {\bibfnamefont {J.~D.}\ \bibnamefont {Tubbs}}, \bibinfo
  {author} {\bibfnamefont {S.~D.}\ \bibnamefont {Kennedy}}, \ and\ \bibinfo
  {author} {\bibfnamefont {D.~H.}\ \bibnamefont {Turner}},\ }\bibfield  {title}
  {\enquote {\bibinfo {title} {Benchmarking {AMBER} force fields for {RNA}:
  Comparisons to {NMR} spectra for single-stranded r({GACC}) are improved by
  revised $\chi$ torsions},}\ }\href@noop {} {\bibfield  {journal} {\bibinfo
  {journal} {J. Phys. Chem. B}\ }\textbf {\bibinfo {volume} {115}},\ \bibinfo
  {pages} {9261--9270} (\bibinfo {year} {2011})}\BibitemShut {NoStop}%
\bibitem [{\citenamefont {Zhao}\ \emph {et~al.}(2020)\citenamefont {Zhao},
  \citenamefont {Kennedy}, \citenamefont {Berger},\ and\ \citenamefont
  {Turner}}]{zhao2020nuclear}%
  \BibitemOpen
  \bibfield  {author} {\bibinfo {author} {\bibfnamefont {J.}~\bibnamefont
  {Zhao}}, \bibinfo {author} {\bibfnamefont {S.~D.}\ \bibnamefont {Kennedy}},
  \bibinfo {author} {\bibfnamefont {K.~D.}\ \bibnamefont {Berger}}, \ and\
  \bibinfo {author} {\bibfnamefont {D.~H.}\ \bibnamefont {Turner}},\ }\bibfield
   {title} {\enquote {\bibinfo {title} {Nuclear magnetic resonance of
  single-stranded {RNA}s and {DNA}s of {CAAU} and {UCAAUC} as benchmarks for
  molecular dynamics simulations},}\ }\href@noop {} {\bibfield  {journal}
  {\bibinfo  {journal} {J. Chem. Theory Comput.}\ }\textbf {\bibinfo {volume}
  {16}},\ \bibinfo {pages} {1968--1984} (\bibinfo {year} {2020})}\BibitemShut
  {NoStop}%
\bibitem [{\citenamefont {Zhao}, \citenamefont {Kennedy},\ and\ \citenamefont
  {Turner}(2022)}]{zhao2022nuclear}%
  \BibitemOpen
  \bibfield  {author} {\bibinfo {author} {\bibfnamefont {J.}~\bibnamefont
  {Zhao}}, \bibinfo {author} {\bibfnamefont {S.~D.}\ \bibnamefont {Kennedy}}, \
  and\ \bibinfo {author} {\bibfnamefont {D.~H.}\ \bibnamefont {Turner}},\
  }\bibfield  {title} {\enquote {\bibinfo {title} {Nuclear magnetic resonance
  spectra and {AMBER} {OL3} and {ROC}-{RNA} simulations of {UCUCGU} reveal
  force field strengths and weaknesses for single-stranded {RNA}},}\
  }\href@noop {} {\bibfield  {journal} {\bibinfo  {journal} {J. Chem. Theory
  Comput.}\ }\textbf {\bibinfo {volume} {18}},\ \bibinfo {pages} {1241--1254}
  (\bibinfo {year} {2022})}\BibitemShut {NoStop}%
\bibitem [{\citenamefont {Ml{\`y}nsk{\`y}}\ \emph {et~al.}(2020)\citenamefont
  {Ml{\`y}nsk{\`y}}, \citenamefont {K{\"u}hrov{\'a}}, \citenamefont {K{\"u}hr},
  \citenamefont {Otyepka}, \citenamefont {Bussi}, \citenamefont
  {Ban{\'a}{\v{s}}},\ and\ \citenamefont {{\v{S}}poner}}]{mlynsky2020fine}%
  \BibitemOpen
  \bibfield  {author} {\bibinfo {author} {\bibfnamefont {V.}~\bibnamefont
  {Ml{\`y}nsk{\`y}}}, \bibinfo {author} {\bibfnamefont {P.}~\bibnamefont
  {K{\"u}hrov{\'a}}}, \bibinfo {author} {\bibfnamefont {T.}~\bibnamefont
  {K{\"u}hr}}, \bibinfo {author} {\bibfnamefont {M.}~\bibnamefont {Otyepka}},
  \bibinfo {author} {\bibfnamefont {G.}~\bibnamefont {Bussi}}, \bibinfo
  {author} {\bibfnamefont {P.}~\bibnamefont {Ban{\'a}{\v{s}}}}, \ and\ \bibinfo
  {author} {\bibfnamefont {J.}~\bibnamefont {{\v{S}}poner}},\ }\bibfield
  {title} {\enquote {\bibinfo {title} {Fine-tuning of the {AMBER} {RNA} force
  field with a new term adjusting interactions of terminal nucleotides},}\
  }\href@noop {} {\bibfield  {journal} {\bibinfo  {journal} {J. Chem. Theory
  Comput.}\ }\textbf {\bibinfo {volume} {16}},\ \bibinfo {pages} {3936--3946}
  (\bibinfo {year} {2020})}\BibitemShut {NoStop}%
\bibitem [{\citenamefont {Bergonzo}\ \emph {et~al.}(2015)\citenamefont
  {Bergonzo}, \citenamefont {Henriksen}, \citenamefont {Roe},\ and\
  \citenamefont {Cheatham}}]{bergonzo2015highly}%
  \BibitemOpen
  \bibfield  {author} {\bibinfo {author} {\bibfnamefont {C.}~\bibnamefont
  {Bergonzo}}, \bibinfo {author} {\bibfnamefont {N.~M.}\ \bibnamefont
  {Henriksen}}, \bibinfo {author} {\bibfnamefont {D.~R.}\ \bibnamefont {Roe}},
  \ and\ \bibinfo {author} {\bibfnamefont {T.~E.}\ \bibnamefont {Cheatham}},\
  }\bibfield  {title} {\enquote {\bibinfo {title} {Highly sampled
  tetranucleotide and tetraloop motifs enable evaluation of common {RNA} force
  fields},}\ }\href@noop {} {\bibfield  {journal} {\bibinfo  {journal} {RNA}\
  }\textbf {\bibinfo {volume} {21}},\ \bibinfo {pages} {1578--1590} (\bibinfo
  {year} {2015})}\BibitemShut {NoStop}%
\bibitem [{\citenamefont {Gil-Ley}, \citenamefont {Bottaro},\ and\
  \citenamefont {Bussi}(2016)}]{gil2016empirical}%
  \BibitemOpen
  \bibfield  {author} {\bibinfo {author} {\bibfnamefont {A.}~\bibnamefont
  {Gil-Ley}}, \bibinfo {author} {\bibfnamefont {S.}~\bibnamefont {Bottaro}}, \
  and\ \bibinfo {author} {\bibfnamefont {G.}~\bibnamefont {Bussi}},\ }\bibfield
   {title} {\enquote {\bibinfo {title} {Empirical corrections to the {AMBER}
  {RNA} force field with target metadynamics},}\ }\href@noop {} {\bibfield
  {journal} {\bibinfo  {journal} {J. Chem. Theory Comput.}\ }\textbf {\bibinfo
  {volume} {12}},\ \bibinfo {pages} {2790--2798} (\bibinfo {year}
  {2016})}\BibitemShut {NoStop}%
\bibitem [{\citenamefont {Aytenfisu}\ \emph {et~al.}(2017)\citenamefont
  {Aytenfisu}, \citenamefont {Spasic}, \citenamefont {Grossfield},
  \citenamefont {Stern},\ and\ \citenamefont {Mathews}}]{Aytenfisu2017}%
  \BibitemOpen
  \bibfield  {author} {\bibinfo {author} {\bibfnamefont {A.~H.}\ \bibnamefont
  {Aytenfisu}}, \bibinfo {author} {\bibfnamefont {A.}~\bibnamefont {Spasic}},
  \bibinfo {author} {\bibfnamefont {A.}~\bibnamefont {Grossfield}}, \bibinfo
  {author} {\bibfnamefont {H.~A.}\ \bibnamefont {Stern}}, \ and\ \bibinfo
  {author} {\bibfnamefont {D.~H.}\ \bibnamefont {Mathews}},\ }\bibfield
  {title} {\enquote {\bibinfo {title} {Revised {RNA} dihedral parameters for
  the {AMBER} force field improve {RNA} molecular dynamics},}\ }\href@noop {}
  {\bibfield  {journal} {\bibinfo  {journal} {J. Chem. Theory Comput.}\
  }\textbf {\bibinfo {volume} {13}},\ \bibinfo {pages} {900--915} (\bibinfo
  {year} {2017})}\BibitemShut {NoStop}%
\bibitem [{\citenamefont {Chen}\ \emph {et~al.}(2022)\citenamefont {Chen},
  \citenamefont {Liu}, \citenamefont {Cui}, \citenamefont {Li},\ and\
  \citenamefont {Chen}}]{chen2022rna}%
  \BibitemOpen
  \bibfield  {author} {\bibinfo {author} {\bibfnamefont {J.}~\bibnamefont
  {Chen}}, \bibinfo {author} {\bibfnamefont {H.}~\bibnamefont {Liu}}, \bibinfo
  {author} {\bibfnamefont {X.}~\bibnamefont {Cui}}, \bibinfo {author}
  {\bibfnamefont {Z.}~\bibnamefont {Li}}, \ and\ \bibinfo {author}
  {\bibfnamefont {H.-F.}\ \bibnamefont {Chen}},\ }\bibfield  {title} {\enquote
  {\bibinfo {title} {{RNA}-specific force field optimization with {CMAP} and
  reweighting},}\ }\href@noop {} {\bibfield  {journal} {\bibinfo  {journal} {J.
  Chem. Inf. Model.}\ }\textbf {\bibinfo {volume} {62}},\ \bibinfo {pages}
  {372--385} (\bibinfo {year} {2022})}\BibitemShut {NoStop}%
\bibitem [{\citenamefont {Yang}\ \emph {et~al.}(2017)\citenamefont {Yang},
  \citenamefont {Lim}, \citenamefont {Kim},\ and\ \citenamefont
  {Pak}}]{Yang2017}%
  \BibitemOpen
  \bibfield  {author} {\bibinfo {author} {\bibfnamefont {C.}~\bibnamefont
  {Yang}}, \bibinfo {author} {\bibfnamefont {M.}~\bibnamefont {Lim}}, \bibinfo
  {author} {\bibfnamefont {E.}~\bibnamefont {Kim}}, \ and\ \bibinfo {author}
  {\bibfnamefont {Y.}~\bibnamefont {Pak}},\ }\bibfield  {title} {\enquote
  {\bibinfo {title} {Predicting {RNA} structures via a simple van der waals
  correction to an all-atom force field},}\ }\href@noop {} {\bibfield
  {journal} {\bibinfo  {journal} {J. Chem. Theory Comput.}\ }\textbf {\bibinfo
  {volume} {13}},\ \bibinfo {pages} {395--399} (\bibinfo {year}
  {2017})}\BibitemShut {NoStop}%
\bibitem [{\citenamefont {Bottaro}\ \emph {et~al.}(2018)\citenamefont
  {Bottaro}, \citenamefont {Bussi}, \citenamefont {Kennedy}, \citenamefont
  {Turner},\ and\ \citenamefont {Lindorff-Larsen}}]{bottaro2018conformational}%
  \BibitemOpen
  \bibfield  {author} {\bibinfo {author} {\bibfnamefont {S.}~\bibnamefont
  {Bottaro}}, \bibinfo {author} {\bibfnamefont {G.}~\bibnamefont {Bussi}},
  \bibinfo {author} {\bibfnamefont {S.~D.}\ \bibnamefont {Kennedy}}, \bibinfo
  {author} {\bibfnamefont {D.~H.}\ \bibnamefont {Turner}}, \ and\ \bibinfo
  {author} {\bibfnamefont {K.}~\bibnamefont {Lindorff-Larsen}},\ }\bibfield
  {title} {\enquote {\bibinfo {title} {Conformational ensembles of {RNA}
  oligonucleotides from integrating {NMR} and molecular simulations},}\
  }\href@noop {} {\bibfield  {journal} {\bibinfo  {journal} {Science Adv.}\
  }\textbf {\bibinfo {volume} {4}},\ \bibinfo {pages} {eaar8521} (\bibinfo
  {year} {2018})}\BibitemShut {NoStop}%
\bibitem [{\citenamefont {Tan}\ \emph {et~al.}(2018)\citenamefont {Tan},
  \citenamefont {Piana}, \citenamefont {Dirks},\ and\ \citenamefont
  {Shaw}}]{tan2018rna}%
  \BibitemOpen
  \bibfield  {author} {\bibinfo {author} {\bibfnamefont {D.}~\bibnamefont
  {Tan}}, \bibinfo {author} {\bibfnamefont {S.}~\bibnamefont {Piana}}, \bibinfo
  {author} {\bibfnamefont {R.~M.}\ \bibnamefont {Dirks}}, \ and\ \bibinfo
  {author} {\bibfnamefont {D.~E.}\ \bibnamefont {Shaw}},\ }\bibfield  {title}
  {\enquote {\bibinfo {title} {{RNA} force field with accuracy comparable to
  state-of-the-art protein force fields},}\ }\href@noop {} {\bibfield
  {journal} {\bibinfo  {journal} {Proc. Natl. Acad. Sci. U.S.A.}\ }\textbf
  {\bibinfo {volume} {115}},\ \bibinfo {pages} {E1346--E1355} (\bibinfo {year}
  {2018})}\BibitemShut {NoStop}%
\bibitem [{\citenamefont {Hecht}(1963)}]{Hecht1963}%
  \BibitemOpen
  \bibfield  {author} {\bibinfo {author} {\bibfnamefont {H.~G.}\ \bibnamefont
  {Hecht}},\ }\bibfield  {title} {\enquote {\bibinfo {title} {Studies of
  delocalized electron bonding},}\ }\href {\doibase 10.1007/BF00529395}
  {\bibfield  {journal} {\bibinfo  {journal} {Theor. Chim. Acta}\ }\textbf
  {\bibinfo {volume} {1}},\ \bibinfo {pages} {133--139} (\bibinfo {year}
  {1963})}\BibitemShut {NoStop}%
\bibitem [{\citenamefont {Bottaro}\ \emph {et~al.}(2019)\citenamefont
  {Bottaro}, \citenamefont {Bussi}, \citenamefont {Pinamonti}, \citenamefont
  {Rei{\ss}er}, \citenamefont {Boomsma},\ and\ \citenamefont
  {Lindorff-Larsen}}]{bottaro2019barnaba}%
  \BibitemOpen
  \bibfield  {author} {\bibinfo {author} {\bibfnamefont {S.}~\bibnamefont
  {Bottaro}}, \bibinfo {author} {\bibfnamefont {G.}~\bibnamefont {Bussi}},
  \bibinfo {author} {\bibfnamefont {G.}~\bibnamefont {Pinamonti}}, \bibinfo
  {author} {\bibfnamefont {S.}~\bibnamefont {Rei{\ss}er}}, \bibinfo {author}
  {\bibfnamefont {W.}~\bibnamefont {Boomsma}}, \ and\ \bibinfo {author}
  {\bibfnamefont {K.}~\bibnamefont {Lindorff-Larsen}},\ }\bibfield  {title}
  {\enquote {\bibinfo {title} {Barnaba: software for analysis of nucleic acid
  structures and trajectories},}\ }\href@noop {} {\bibfield  {journal}
  {\bibinfo  {journal} {RNA}\ }\textbf {\bibinfo {volume} {25}},\ \bibinfo
  {pages} {219--231} (\bibinfo {year} {2019})}\BibitemShut {NoStop}%
\bibitem [{\citenamefont {Davies}(1978)}]{Davies1978}%
  \BibitemOpen
  \bibfield  {author} {\bibinfo {author} {\bibfnamefont {D.~B.}\ \bibnamefont
  {Davies}},\ }\bibfield  {title} {\enquote {\bibinfo {title} {Conformations of
  nucleosides and nucleotides},}\ }\href {\doibase
  https://doi.org/10.1016/0079-6565(78)80006-5} {\bibfield  {journal} {\bibinfo
   {journal} {Prog. Nucl. Magn. Reson. Spectrosc.}\ }\textbf {\bibinfo {volume}
  {12}},\ \bibinfo {pages} {135--225} (\bibinfo {year} {1978})}\BibitemShut
  {NoStop}%
\bibitem [{\citenamefont {Lankhorst}\ \emph {et~al.}(1985)\citenamefont
  {Lankhorst}, \citenamefont {Haasnoot}, \citenamefont {Erkelens},
  \citenamefont {Westerink}, \citenamefont {Marel}, \citenamefont {Boom},\ and\
  \citenamefont {Altona}}]{Lankhorst1985}%
  \BibitemOpen
  \bibfield  {author} {\bibinfo {author} {\bibfnamefont {P.}~\bibnamefont
  {Lankhorst}}, \bibinfo {author} {\bibfnamefont {C.~A.~G.}\ \bibnamefont
  {Haasnoot}}, \bibinfo {author} {\bibfnamefont {C.}~\bibnamefont {Erkelens}},
  \bibinfo {author} {\bibfnamefont {H.}~\bibnamefont {Westerink}}, \bibinfo
  {author} {\bibfnamefont {G.}~\bibnamefont {Marel}}, \bibinfo {author}
  {\bibfnamefont {J.}~\bibnamefont {Boom}}, \ and\ \bibinfo {author}
  {\bibfnamefont {C.}~\bibnamefont {Altona}},\ }\bibfield  {title} {\enquote
  {\bibinfo {title} {Carbon-13 {NMR} in conformational analysis of nucleic acid
  fragments. 4. the torsion angle distribution about the {C}3'-{O}3' bond in
  {DNA} constituents},}\ }\href {\doibase 10.1093/nar/13.3.927} {\bibfield
  {journal} {\bibinfo  {journal} {Nucleic Acids Res.}\ }\textbf {\bibinfo
  {volume} {13}},\ \bibinfo {pages} {927--42} (\bibinfo {year}
  {1985})}\BibitemShut {NoStop}%
\bibitem [{\citenamefont {Mooren}\ \emph {et~al.}(1994)\citenamefont {Mooren},
  \citenamefont {Wijmenga}, \citenamefont {van~der Marel}, \citenamefont {van
  Boom},\ and\ \citenamefont {Hilbers}}]{mooren1994solution}%
  \BibitemOpen
  \bibfield  {author} {\bibinfo {author} {\bibfnamefont {M.~M.}\ \bibnamefont
  {Mooren}}, \bibinfo {author} {\bibfnamefont {S.~S.}\ \bibnamefont
  {Wijmenga}}, \bibinfo {author} {\bibfnamefont {G.~A.}\ \bibnamefont {van~der
  Marel}}, \bibinfo {author} {\bibfnamefont {J.~H.}\ \bibnamefont {van Boom}},
  \ and\ \bibinfo {author} {\bibfnamefont {C.~W.}\ \bibnamefont {Hilbers}},\
  }\bibfield  {title} {\enquote {\bibinfo {title} {The solution structure of
  the circular trinucleotide cr ({GpGpGp}) determined by {NMR} and molecular
  mechanics calculation},}\ }\href@noop {} {\bibfield  {journal} {\bibinfo
  {journal} {Nucleic Acids Res.}\ }\textbf {\bibinfo {volume} {22}},\ \bibinfo
  {pages} {2658--2666} (\bibinfo {year} {1994})}\BibitemShut {NoStop}%
\bibitem [{\citenamefont {Lee}\ and\ \citenamefont
  {Sarma}(1976)}]{lee1976aqueous}%
  \BibitemOpen
  \bibfield  {author} {\bibinfo {author} {\bibfnamefont {C.-H.}\ \bibnamefont
  {Lee}}\ and\ \bibinfo {author} {\bibfnamefont {R.~H.}\ \bibnamefont
  {Sarma}},\ }\bibfield  {title} {\enquote {\bibinfo {title} {Aqueous solution
  conformation of rigid nucleosides and nucleotides},}\ }\href@noop {}
  {\bibfield  {journal} {\bibinfo  {journal} {J. Am. Chem. Soc.}\ }\textbf
  {\bibinfo {volume} {98}},\ \bibinfo {pages} {3541--3548} (\bibinfo {year}
  {1976})}\BibitemShut {NoStop}%
\bibitem [{\citenamefont {Blackburn}, \citenamefont {Lapper},\ and\
  \citenamefont {Smith}(1973)}]{blackburn1973proton}%
  \BibitemOpen
  \bibfield  {author} {\bibinfo {author} {\bibfnamefont {B.~J.}\ \bibnamefont
  {Blackburn}}, \bibinfo {author} {\bibfnamefont {R.~D.}\ \bibnamefont
  {Lapper}}, \ and\ \bibinfo {author} {\bibfnamefont {I.~C.}\ \bibnamefont
  {Smith}},\ }\bibfield  {title} {\enquote {\bibinfo {title} {Proton magnetic
  resonance study of the conformations of 3', 5'-cyclic nucleotides},}\
  }\href@noop {} {\bibfield  {journal} {\bibinfo  {journal} {J. Am. Chem.
  Soc.}\ }\textbf {\bibinfo {volume} {95}},\ \bibinfo {pages} {2873--2878}
  (\bibinfo {year} {1973})}\BibitemShut {NoStop}%
\bibitem [{\citenamefont {Altona}(1982)}]{altona1982conformational}%
  \BibitemOpen
  \bibfield  {author} {\bibinfo {author} {\bibfnamefont {C.}~\bibnamefont
  {Altona}},\ }\bibfield  {title} {\enquote {\bibinfo {title} {Conformational
  analysis of nucleic acids. determination of backbone geometry of
  single-helical {RNA} and {DNA} in aqueous solution},}\ }\href@noop {}
  {\bibfield  {journal} {\bibinfo  {journal} {Recl. Trav. Chim. Pays-Bas}\
  }\textbf {\bibinfo {volume} {101}},\ \bibinfo {pages} {413--433} (\bibinfo
  {year} {1982})}\BibitemShut {NoStop}%
\bibitem [{\citenamefont {Lindorff-Larsen}, \citenamefont {Best},\ and\
  \citenamefont {Vendruscolo}(2005)}]{lindorff2005interpreting}%
  \BibitemOpen
  \bibfield  {author} {\bibinfo {author} {\bibfnamefont {K.}~\bibnamefont
  {Lindorff-Larsen}}, \bibinfo {author} {\bibfnamefont {R.~B.}\ \bibnamefont
  {Best}}, \ and\ \bibinfo {author} {\bibfnamefont {M.}~\bibnamefont
  {Vendruscolo}},\ }\bibfield  {title} {\enquote {\bibinfo {title}
  {Interpreting dynamically-averaged scalar couplings in proteins},}\
  }\href@noop {} {\bibfield  {journal} {\bibinfo  {journal} {J. Biomol. NMR}\
  }\textbf {\bibinfo {volume} {32}},\ \bibinfo {pages} {273--280} (\bibinfo
  {year} {2005})}\BibitemShut {NoStop}%
\bibitem [{\citenamefont {Shen}\ and\ \citenamefont
  {Hamelberg}(2008)}]{shen2008statistical}%
  \BibitemOpen
  \bibfield  {author} {\bibinfo {author} {\bibfnamefont {T.}~\bibnamefont
  {Shen}}\ and\ \bibinfo {author} {\bibfnamefont {D.}~\bibnamefont
  {Hamelberg}},\ }\bibfield  {title} {\enquote {\bibinfo {title} {A statistical
  analysis of the precision of reweighting-based simulations},}\ }\href@noop {}
  {\bibfield  {journal} {\bibinfo  {journal} {J. Chem. Phys.}\ }\textbf
  {\bibinfo {volume} {129}},\ \bibinfo {pages} {034103} (\bibinfo {year}
  {2008})}\BibitemShut {NoStop}%
\bibitem [{\citenamefont {Rangan}\ \emph {et~al.}(2018)\citenamefont {Rangan},
  \citenamefont {Bonomi}, \citenamefont {Heller}, \citenamefont {Cesari},
  \citenamefont {Bussi},\ and\ \citenamefont
  {Vendruscolo}}]{rangan2018determination}%
  \BibitemOpen
  \bibfield  {author} {\bibinfo {author} {\bibfnamefont {R.}~\bibnamefont
  {Rangan}}, \bibinfo {author} {\bibfnamefont {M.}~\bibnamefont {Bonomi}},
  \bibinfo {author} {\bibfnamefont {G.~T.}\ \bibnamefont {Heller}}, \bibinfo
  {author} {\bibfnamefont {A.}~\bibnamefont {Cesari}}, \bibinfo {author}
  {\bibfnamefont {G.}~\bibnamefont {Bussi}}, \ and\ \bibinfo {author}
  {\bibfnamefont {M.}~\bibnamefont {Vendruscolo}},\ }\bibfield  {title}
  {\enquote {\bibinfo {title} {Determination of structural ensembles of
  proteins: restraining vs reweighting},}\ }\href@noop {} {\bibfield  {journal}
  {\bibinfo  {journal} {J. Chem. Theory Comput.}\ }\textbf {\bibinfo {volume}
  {14}},\ \bibinfo {pages} {6632--6641} (\bibinfo {year} {2018})}\BibitemShut
  {NoStop}%
\bibitem [{\citenamefont {Sanjabi}\ \emph {et~al.}(2018)\citenamefont
  {Sanjabi}, \citenamefont {Ba}, \citenamefont {Razaviyayn},\ and\
  \citenamefont {Lee}}]{sanjabi2018convergence}%
  \BibitemOpen
  \bibfield  {author} {\bibinfo {author} {\bibfnamefont {M.}~\bibnamefont
  {Sanjabi}}, \bibinfo {author} {\bibfnamefont {J.}~\bibnamefont {Ba}},
  \bibinfo {author} {\bibfnamefont {M.}~\bibnamefont {Razaviyayn}}, \ and\
  \bibinfo {author} {\bibfnamefont {J.~D.}\ \bibnamefont {Lee}},\ }\bibfield
  {title} {\enquote {\bibinfo {title} {On the convergence and robustness of
  training gans with regularized optimal transport},}\ }\href@noop {}
  {\bibfield  {journal} {\bibinfo  {journal} {Advances in Neural Information
  Processing Systems}\ }\textbf {\bibinfo {volume} {31}} (\bibinfo {year}
  {2018})}\BibitemShut {NoStop}%
\bibitem [{\citenamefont {Herrmann}, \citenamefont {G{\"u}ntert},\ and\
  \citenamefont {W{\"u}thrich}(2002)}]{herrmann2002protein}%
  \BibitemOpen
  \bibfield  {author} {\bibinfo {author} {\bibfnamefont {T.}~\bibnamefont
  {Herrmann}}, \bibinfo {author} {\bibfnamefont {P.}~\bibnamefont
  {G{\"u}ntert}}, \ and\ \bibinfo {author} {\bibfnamefont {K.}~\bibnamefont
  {W{\"u}thrich}},\ }\bibfield  {title} {\enquote {\bibinfo {title} {Protein
  {NMR} structure determination with automated {NOE} assignment using the new
  software {CANDID} and the torsion angle dynamics algorithm {DYANA}},}\
  }\href@noop {} {\bibfield  {journal} {\bibinfo  {journal} {J. Mol. Biol.}\
  }\textbf {\bibinfo {volume} {319}},\ \bibinfo {pages} {209--227} (\bibinfo
  {year} {2002})}\BibitemShut {NoStop}%
\bibitem [{\citenamefont {Lange}(2014)}]{lange2014automatic}%
  \BibitemOpen
  \bibfield  {author} {\bibinfo {author} {\bibfnamefont {O.~F.}\ \bibnamefont
  {Lange}},\ }\bibfield  {title} {\enquote {\bibinfo {title} {Automatic {NOESY}
  assignment in {CS-RASREC-Rosetta}},}\ }\href@noop {} {\bibfield  {journal}
  {\bibinfo  {journal} {J. Biomol. NMR}\ }\textbf {\bibinfo {volume} {59}},\
  \bibinfo {pages} {147--159} (\bibinfo {year} {2014})}\BibitemShut {NoStop}%
\bibitem [{\citenamefont {Pesce}\ and\ \citenamefont
  {Lindorff-Larsen}(2021)}]{pesce2021refining}%
  \BibitemOpen
  \bibfield  {author} {\bibinfo {author} {\bibfnamefont {F.}~\bibnamefont
  {Pesce}}\ and\ \bibinfo {author} {\bibfnamefont {K.}~\bibnamefont
  {Lindorff-Larsen}},\ }\bibfield  {title} {\enquote {\bibinfo {title}
  {Refining conformational ensembles of flexible proteins against small-angle
  {X}-ray scattering data},}\ }\href@noop {} {\bibfield  {journal} {\bibinfo
  {journal} {Biophy. J.}\ }\textbf {\bibinfo {volume} {120}},\ \bibinfo {pages}
  {5124--5135} (\bibinfo {year} {2021})}\BibitemShut {NoStop}%
\bibitem [{\citenamefont {Bernetti}, \citenamefont {Hall},\ and\ \citenamefont
  {Bussi}(2021)}]{bernetti2021reweighting}%
  \BibitemOpen
  \bibfield  {author} {\bibinfo {author} {\bibfnamefont {M.}~\bibnamefont
  {Bernetti}}, \bibinfo {author} {\bibfnamefont {K.~B.}\ \bibnamefont {Hall}},
  \ and\ \bibinfo {author} {\bibfnamefont {G.}~\bibnamefont {Bussi}},\
  }\bibfield  {title} {\enquote {\bibinfo {title} {Reweighting of molecular
  simulations with explicit-solvent {SAXS} restraints elucidates ion-dependent
  {RNA} ensembles},}\ }\href@noop {} {\bibfield  {journal} {\bibinfo  {journal}
  {Nucleic Acids Res.}\ }\textbf {\bibinfo {volume} {49}},\ \bibinfo {pages}
  {e84--e84} (\bibinfo {year} {2021})}\BibitemShut {NoStop}%
\end{thebibliography}%

\end{document}